\documentclass[12pt,journal,final,onecolumn]{IEEEtran}

\usepackage[dvips]{graphicx}
\usepackage{epsfig}
\usepackage[cmex10]{amsmath}
\usepackage{amssymb}
\usepackage{amsthm}
\usepackage{amsfonts}
\usepackage{bm}
\usepackage{dsfont}
\usepackage{color}
\usepackage[mathscr]{eucal}
\usepackage{cite}
\usepackage{color}

\graphicspath{{figs/}}

\input{niesen.def}

\begin{document}

\bibliographystyle{ieeetr}

\title{The Approximate Capacity of 
\\ the Gaussian $N$-Relay Diamond Network} 

\author{Urs~Niesen and Suhas~N.~Diggavi%
\thanks{U. Niesen is with the
Mathematics of Networks and Communications Research Department, Bell
Labs, Alcatel-Lucent.
Email: urs.niesen@alcatel-lucent.com.
S. Diggavi is with the University of California, Los Angeles.
Email: suhas@ee.ucla.edu.}%
}

\maketitle

\begin{abstract} 
    We consider the Gaussian ``diamond'' or parallel relay network, in
    which a source node transmits a message to a destination node with
    the help of $N$ relays. Even for the symmetric setting, in
    which the channel gains to the relays are identical and the channel
    gains from the relays are identical, the capacity of this channel is
    unknown in general. The best known capacity approximation is up to
    an additive gap of order $N$ bits and up to a multiplicative gap
    of order $N^2$, with both gaps independent of the channel gains.

    In this paper, we approximate the capacity of the symmetric Gaussian
    $N$-relay diamond network up to an additive gap of $1.8$ bits and up
    to a multiplicative gap of a factor $14$. Both gaps are independent
    of the channel gains and, unlike the best previously known result,
    are also independent of the number of relays $N$ in the network.
    Achievability is based on bursty amplify-and-forward, showing that
    this simple scheme is uniformly approximately optimal, both in the
    low-rate as well as in the high-rate regimes. The upper bound on
    capacity is based on a careful evaluation of the cut-set bound. We
    also present approximation results for the asymmetric Gaussian
    $N$-relay diamond network. In particular, we show that bursty
    amplify-and-forward combined with optimal relay selection achieves a
    rate within a factor $O(\log^4(N))$ of capacity with pre-constant in
    the order notation independent of the channel gains. 
\end{abstract}

\section{Introduction}
\label{sec:intro}

\begin{figure}[htbp]
    \begin{center}
        \scalebox{1}{\hspace{1.6cm}\input{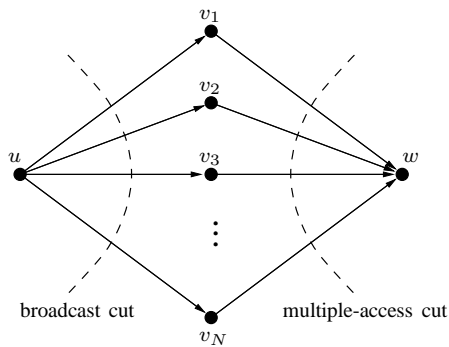}} 
    \end{center}

    \caption{The $N$-relay diamond network. The source node $u$
    transmits a message to the destination node $w$ via the $N$ relays
    $\{v_n\}_{n=1}^N$. The two cuts indicated in the figure are the
    broadcast cut (separating the source $u$ from the relays $\{v_n\}$)
    and the multiple-access cut (separating the relays $\{v_n\}$ from
    the destination $w$).}

    \label{fig:diamond_cuts1}
\end{figure}

Cooperation is a key feature of wireless communication. A simple
canonical channel model capturing this feature is the ``diamond'' or
parallel relay network introduced by Schein and
Gallager \cite{schein00, schein01}. This network consists of a source node
connected through a broadcast channel to $N$ relays; the relays, in
turn, are connected to the destination node through a multiple-access
channel (see Fig.~\ref{fig:diamond_cuts1}).
The objective is to maximize the rate achievable between the source and
the destination with the help of the $N$ relays. Throughout this paper,
we will be interested in the Gaussian version of this problem, in which
both the broadcast and the multiple-access parts are subject to additive
Gaussian noise. Moreover, for simplicity we will restrict attention in a
significant part of the paper to the symmetric case, in which the
channel gains within the multiple-access part and within the broadcast
part of the network are identical (but are allowed to differ between the
multiple-access and broadcast parts). However, we do show that some of
the results for the symmetric setting can be extended to the asymmetric
setting.

For the Gaussian $2$-relay diamond network, the rates achievable with
decode-and-forward and with amplify-and-forward at the relays were
analyzed in \cite{schein01}. It is shown there that these schemes achieve
capacity in some regimes of signal-to-noise ratios (SNRs) of the
broadcast and multiple-access parts of the diamond network. The
asymptotic behavior of the $N$-relay Gaussian diamond network was
investigated in \cite{gastpar05}. In certain regimes of SNRs of the
broadcast and multiple-access parts of the network, it is shown that
amplify-and-forward is capacity achieving in the limit as $N\to\infty$.
New achievable schemes for the Gaussian diamond network with bandwidth
mismatch (i.e., the source and the relays have different bandwidth) were
introduced in \cite{kochman08} and \cite{rezaei09}. Perhaps
surprisingly, these schemes lead to higher achievable rates than the
ones obtained with amplify-and-forward and decode-and-forward even when
the bandwidths at the source and the relays are identical. Half-duplex
versions of the Gaussian diamond network, in which the relays cannot
receive and transmit signals simultaneously, were considered in
\cite{xue07} and \cite{bagheri09}. The capacity of a special class of
$2$-relay diamond networks is derived in \cite{kang11}. For networks in
this class, one relay receives the signal sent at the source without
noise, and the destination node is connected to the relays by two
orthogonal bit pipes of fixed rate. To the best of our knowledge, this
is the only non-trivial example for which the capacity of the diamond
network is known for all values of SNR. For the general Gaussian
$N$-relay diamond network, the capacity is unknown.

Given the difficulty of determining the capacity of communication
networks in general and of the diamond network in particular, it is
natural to ask if it can at least be approximated. For high rates, such
an approximation should be additive in nature, i.e., we would like to
determine capacity up to an additive gap. For low rates, such an
approximation should be multiplicative, i.e., we would like to determine
capacity up to a multiplicative gap. If a communication strategy can be
shown to have both small additive as well as multiplicative gaps, then
this strategy is provably close to optimal both in the high rate as well
as low rate regimes.

Additive approximations for channel capacity of communication networks
were first derived in \cite{etkin08}, where the capacity region of the
two-user Gaussian interference channel is determined up to an additive
gap of one bit. This was mainly enabled through a new outer bound for
the interference channel. The approach of approximate capacity
characterization was applied to general relay networks with
single-source multicast in \cite{avestimehr09}. By introducing a new
relaying strategy termed quantize-map-forward, capacity is derived up to
an additive gap of $15n$ bits, where $n$ is the number of nodes in the
network. This additive gap was improved through the use of vector
quantization at the relays \cite{ozgur10c,lim11}. The sharpest known
additive approximation gap is $1.26n$ bits for the complex Gaussian case
(or $0.63n$ for the real case) \cite{lim11}. Since the $N$-user diamond
network is a special case of a relay network with a single source and
destination and with $n=N+2$ nodes, these results yield an additive
approximation up to a gap of $0.63N+1.26$ bits for this network
(assuming real channel gains).

Multiplicative approximations were mostly analyzed for large wireless
networks, for which the rate per source-destination pair is low.  For a
network with $n$ nodes, the emphasis is on finding capacity
approximations up to a small multiplicative factor in $n$. This approach
was pioneered in \cite{gupta00a}. Under a restricted model of
communication, (essentially) the equal rate point of the capacity region
of a wireless network with $n$ randomly placed nodes was determined up
to a constant multiplicative factor independent of $n$. Without the
restrictive communication assumptions in \cite{gupta00a}, the problem
becomes considerably harder. Approximations for the equal rate point
under a Gaussian model were derived in \cite{ozgur07b} up to a
multiplicative factor of $O(n^{\varepsilon})$ for any $\varepsilon > 0$.
These approximation results were subsequently sharpened in
\cite{ghaderi09, niesen09a} to a factor $n^{O(1/\sqrt{\log(n)})}$.
Under some conditions on the node placement, this factor can further be
sharpened to $O(\log(n))$ \cite{niesen09c}. Multiplicative
approximations for arbitrary relay networks with single-source multicast
(as opposed to wireless networks with multiple unicast, i.e., multiple
separate source-destination pairs) were derived in \cite{avestimehr09}.
For a network with maximum degree $d$, the capacity is approximated to
within a factor of $2d(d+1)$. As pointed out earlier, the Gaussian
$N$-relay diamond network is such a network with maximum degree $d=N$,
and hence this result yields a multiplicative approximation up to a
factor of $2N(N+1)$.

To summarize, the capacity region of the general Gaussian $N$-relay
diamond network is not known. The best known additive approximation is
up to a gap of $0.63N+1.26$ bits, and the best known multiplicative
approximation is up to a factor of $2N(N+1)$. In either case, the bounds
degrade rather quickly as $N$ increases. It is hence of interest to find
approximation guarantees that behave better as a function of the number
of relays $N$ in the network. Ideally, we would like the approximation
guarantees to be uniform in in the network size. 

As a main result of this paper, we show that such a uniform
approximation is indeed possible. More precisely, we find an additive
approximation of the capacity of the symmetric Gaussian $N$-relay
diamond network of gap at most $1.8$ bits for any SNR and number of
relays $N$. Moreover, we find a multiplicative approximation to the
capacity up to at most a factor $14$, again for any SNR and number of
relays $N$. This is a significant improvement over the previously best
known additive approximation of $0.63N+1.26$ bits and multiplicative
approximation of a factor $2N(N+1)$, especially for large values of $N$.
In particular, as far as we know, this is the first such approximation
result (both multiplicative as well as additive) that is independent of
the number of network nodes for a nontrivial class of wireless networks.  

We further show that bursty amplify-and-forward (first introduced in
\cite[p.~76]{schein01}) with properly chosen duty cycle is close to
capacity achieving for the diamond network simultaneously in the sense
of multiplicative and additive approximation up to the aforementioned
gaps. Hence, bursty amplify-and-forward with appropriately chosen duty
cycle is a good communication scheme for the symmetric Gaussian
$N$-relay diamond network both at low and at high SNRs, and
independently of the number of relays $N$. 

Some of these results can be extended to the asymmetric setting.
For general (i.e., not necessarily symmetric) Gaussian $N$-relay diamond
networks, we provide a factor $O(\log^4(N))$ multiplicative
approximation of capacity, with pre-constant in the order notation
independent of the channel gains. Achievability is based again on bursty
amplify-and-forward, but this time a careful selection of relays is also
necessary.

The main technical contribution of this paper is the upper bound on
capacity. The standard way to obtain upper bounds on the capacity of the
diamond network is to evaluate two particular cuts in the wireless
network, namely the one separating the source from the relays (called
the \emph{broadcast cut} in the following) and the one separating the
relays from the destination (called the \emph{multiple-access cut} in
the following) as depicted in Fig.~\ref{fig:diamond_cuts1}. This
approach is taken, for example, in \cite{gastpar05, kochman08,
rezaei09}. In fact, for symmetric Gaussian $N$-relay diamond networks,
whenever the capacity is known, it coincides with the minimum of these
two cuts. We show in this paper that, in order to obtain uniform
additive or multiplicative approximations for the capacity of this
network, considering just these two cuts is not sufficient. Instead we
need to \emph{simultaneously} optimize over \emph{all} possible $2^N$
cuts separating the source from the destination. Without this careful
outer bound evaluation, we believe that the uniform (in network size)
approximation would not have been possible.

The remainder of this paper is organized as follows.
Section~\ref{sec:problem} formally introduces the problem statement.
Section~\ref{sec:main} presents the main results; the corresponding
proofs are presented in Section~\ref{sec:proofs}.
Section~\ref{sec:conclusion} contains concluding remarks.

\section{Problem Statement}
\label{sec:problem}

\begin{figure}[htbp]
    \begin{center}
        \scalebox{1}{\input{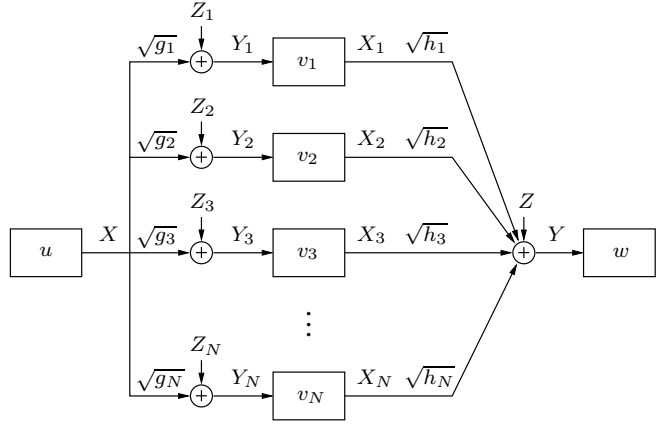}}
    \end{center}

    \caption{The Gaussian $N$-relay diamond network.}

    \label{fig:diamond}
\end{figure}

Consider the Gaussian $N$-relay diamond network as depicted in
Fig.~\ref{fig:diamond}. 
The source node $u$ transmits a message to the destination node $w$ with
the help of $N$ parallel relays $\{v_1,\ldots,v_N\}$. The channel inputs
at time $t\in\N$ at nodes $u$ and $v_n$ are denoted by $X[t]$ and
$X_n[t]$, respectively. The channel outputs at time $t\in\N$ at nodes
$w$ and $v_n$ are denoted by $Y[t]$ and $Y_n[t]$. The channel inputs
and outputs are related as
\begin{align*}
    Y_n[t] & \defeq \sqrt{g_n}X[t]+Z_n[t], \\ 
    Y[t] & \defeq \sum_{n=1}^N {\textstyle \sqrt{h_n}} X_n[t]+Z[t],
\end{align*}
where $(Z[t])_{t}, (Z_n[t])_{n,t}$ are independent and identically
distributed Gaussian random variables with mean zero and variance one,
independent of the channel inputs. The channel gains $(g_n)_{n=1}^N$ and
$(h_n)_{n=1}^N$ are assumed to be real positive numbers, constant as a
function of time, and known throughout the network.

A \emph{$T$-length block code} for the diamond network is a
collection of functions 
\begin{align*}
    f\from & \{1,\ldots, M\} \to \R^T, \\
    f_n\from & \R^T \to \R^T, \ \forall n\in\{1,\ldots, N\}, \\
    \phi\from & \R^T\to \{1,\ldots, M\}.
\end{align*}
The encoding function $f$ maps the message $W$,
assumed to be uniformly distributed over the set $\{1,\ldots, M\}$, to 
the channel inputs
\begin{equation*}
    (X[t])_{t=1}^T\defeq f(W)
\end{equation*}
at the source node $u$. The function $f_n$ maps the channel outputs
$(Y_n[t])_{t=1}^T$ to the channel inputs
\begin{equation*}
    (X_n[t])_{t=1}^T\defeq f_n\big((Y_n[t])_{t=1}^T\big)
\end{equation*}
at relay $v_n$.\footnote{Note that the functions $f_n$ at the relays are
not causal. This is to simplify notation; due to the layered
nature of the network all results remain the same if causality is
imposed.} The decoding function $\phi$ maps the
channel outputs $(Y[t])_{t=1}^T$ at the destination node $w$ into a
reconstruction
\begin{equation*}
    \hat{W} \defeq \phi\big((Y[t])_{t=1}^T\big).
\end{equation*}

We say the code satisfies a \emph{unit average power constraint} if 
\begin{align*}
    \frac{1}{T}\sum_{t=1}^T \E(X^2[t]) & \leq 1, \\
    \frac{1}{T}\sum_{t=1}^T \E(X_n^2[t]) & \leq 1, \ \forall n\in\{1,\ldots, N\}.
\end{align*}
The \emph{rate} of the code is
\begin{equation*}
    \log(M)/T,
\end{equation*}
and its \emph{average probability of error}
\begin{equation*}
    \Pp(\hat{W}\neq W).    
\end{equation*}
A rate $R$ is \emph{achievable} if there exists a sequence of $T$-length
block codes with unit average power constraint and rate at least $R$
such that the average probability of error approaches zero as
$T\to\infty$. The \emph{capacity} $C\big(N,(g_n),(h_n)\big)$ of the
diamond network is the supremum of all achievable rates.

A natural scheme for the diamond network is \emph{amplify-and-forward},
in which each relay transmits a scaled version of the received signal.
Formally, 
\begin{equation*}
    X_n[t] 
    = \alpha_n Y_n[t]
    = \alpha_n \sqrt{g_n}X[t] + \alpha_n Z_n[t],
\end{equation*}
where the constant $\alpha_n$ is chosen to satisfy the power constraint
at the relay. Denote by $R_1\big(N,(g_n),(h_n)\big)$ the rate achieved
by amplify-and-forward with optimal choice of $(\alpha_n)_{n=1}^N$.  We
point out that the optimization over $(\alpha_n)_{n=1}^N$ is necessary.
While perhaps counterintuitive, it turns out that in the asymmetric
setting the rate of amplify-and-forward is not always maximized when the
relays use all available transmit power (see \cite{schein01} for a
discussion of this phenomenon in the two-relay case). 

If the SNR at the relays is low (i.e., $g_n\ll 1$), it can be shown that
simple amplify-and-forward is arbitrarily suboptimal. This is because
the received signal power $g_n$ at the relay $v_n$ is much smaller than
the noise power $1$, and therefore the relay amplifies mostly noise.
This effect can be mitigated by using \emph{bursty amplify-and-forward}
\cite{schein01}. For a constant $\delta\in(0,1]$, called the \emph{duty
cycle} in the following, we communicate for a fraction $\delta$ of time
at average power $1/\delta$ using the amplify-and-forward scheme and
stay silent for the remaining time. This satisfies the overall average
unit power constraint. The resulting achievable rate is denoted by
$R_\delta\big(N,(g_n),(h_n)\big)$. This notation is consistent,
i.e., for $\delta=1$ the simple and bursty amplify-and-forward schemes
coincide and achieve both rate $R_1\big(N,(g_n),(h_n)\big)$.

A special case of the general diamond network described so far is the
\emph{symmetric} setting, in which $g_1=g_2=\ldots = g_N = g$ and
$h_1=h_2\ldots = h_N = h$. With slight abuse of notation, we denote the
capacity and rates achievable by bursty amplify-and-forward for the
symmetric setting by $C(N,g,h)$ and $R_\delta(N,g,h)$.

Throughout this paper, we use bold font to denote vectors and matrices.
$\log(\cdot)$ and $\ln(\cdot)$ denote the logarithms to base $2$ and
$e$, respectively. All capacities and rates are expressed in bits per
channel use.

\section{Main Results}
\label{sec:main}

The main results of this paper are additive and multiplicative capacity
approximations for the Gaussian diamond relay network. We start with a
discussion of symmetric networks in Section~\ref{sec:main_symmetric}.
General asymmetric networks are treated in
Section~\ref{sec:main_asymmetric}.

\subsection{Symmetric Diamond Networks}
\label{sec:main_symmetric}

The first result lower bounds the rate achievable over a symmetric
diamond network by using bursty amplify-and-forward with optimized duty
cycle $\delta$.
\begin{theorem}
    \label{thm:achievability}
    For every symmetric diamond network with  $N\geq 2$ relays and
    channel gains $g,h> 0$, there exists a duty cycle
    $\delta^\star\in(0,1]$ such that bursty amplify-and-forward achieves
    at least the rate 
    \begin{equation*}
        R_{\delta^\star}(N,g,h)
        \geq
        \begin{cases}
            \tfrac{1}{2}\log\big(1+\tfrac{1}{3}N\min\{g,Nh\}\big),
            & \text{if $\max\{g,Nh\} \geq 1$} \\
            \tfrac{1}{2}\ln(4/3)\log(1+Ng),
            & \text{if $\max\{g,Nh\} < 1$, $g \leq h$} \\
            \tfrac{1}{2}\log\big(1+\tfrac{1}{3}N^2gh\big),
            & \text{if $\max\{g,Nh\} < 1$, $g \in(h,N^2h)$, $N\sqrt{gh}\geq 1$} \\
            \tfrac{1}{2}\ln(4/3)\log(1+N\sqrt{gh}),
            & \text{if $\max\{g,Nh\} < 1$, $g \in(h,N^2h)$, $N\sqrt{gh} < 1$} \\
            \tfrac{1}{2}\ln(4/3)\log(1+N^2h),
            & \text{if $\max\{g,Nh\} < 1$, $g \geq N^2h$}.
        \end{cases}
    \end{equation*}
\end{theorem}

The proof of Theorem~\ref{thm:achievability} is presented in
Section~\ref{sec:proofs_achievability}.  Note that the optimal duty
cycle $\delta^\star$ is allowed to depend on $N$, $g$, and $h$. In the
high-rate regime, i.e., the first and third cases in
Theorem~\ref{thm:achievability}, the duty cycle achieving the lower
bound is $\delta^\star = 1$, and hence the bursty amplify-and-forward
scheme reduces to simple amplify-and-forward.  On the other hand, in the
low-rate regime, i.e., the second, fourth, and fifth cases in
Theorem~\ref{thm:achievability}, $\delta^\star < 1$, and (genuine)
bursty amplify-and-forward is used.

Having established an achievable rate, the next theorem provides an
upper bound on the capacity of the diamond network.
\begin{theorem}
    \label{thm:converse}
    For every symmetric diamond network with $N\geq 2$ relays and
    channel gains $g,h> 0$, capacity is upper bounded by
    \begin{equation*}
        C(N,g,h)
        \leq
        \begin{cases}
            \tfrac{1}{2}\log\big(1+N\min\{g,Nh\}\big),
            & \text{if $\max\{g,Nh\} \geq 1$} \\
            \tfrac{1}{2}\log(1+Ng),
            & \text{if $\max\{g,Nh\} < 1$, $g \leq h$} \\
            \tfrac{1}{2}\log\big(1+2N^2gh\big)+\tfrac{1}{2},
            & \text{if $\max\{g,Nh\} < 1$, $g \in(h,N^2h)$, $N\sqrt{gh}\geq 1$} \\
            \log(1+2N\sqrt{gh}),
            & \text{if $\max\{g,Nh\} < 1$, $g \in(h,N^2h)$, $N\sqrt{gh} < 1$} \\
            \tfrac{1}{2}\log(1+N^2h),
            & \text{if $\max\{g,Nh\} < 1$, $g \geq N^2h$}.
        \end{cases}
    \end{equation*}
\end{theorem}

The proof of Theorem~\ref{thm:converse} is presented in
Section~\ref{sec:proofs_converse}.  As a corollary to
Theorems~\ref{thm:achievability} and \ref{thm:converse}, we obtain that
bursty amplify-and-forward is close to optimal, in the sense that it
achieves capacity both up to a constant additive gap as well as a
constant multiplicative gap, where both constants are independent of the
number of relays $N$ and the channel gains $g$ and $h$. This shows that
optimized bursty amplify-and-forward is a good communication scheme for
the symmetric diamond network both at low rates (due to the small
multiplicative gap) as well as at high rates (due to the small additive
gap).

\begin{samepage}
    \begin{corollary}
        \label{thm:approx}
        For every symmetric diamond network with $N\geq 2$ relays and
        channel gains $g,h> 0$, there exists a duty cycle
        $\delta^\star\in(0,1]$ such that
        \begin{equation*}
            C(N,g,h)-R_{\delta^\star}(N,g,h)
            \leq 1+\tfrac{1}{2}\log(3)
            \leq 1.8 \text{ bits}, 
        \end{equation*} 
        \nopagebreak[4] and 
        \begin{equation*}
            \frac{C(N,g,h)}{R_{\delta^\star}(N,g,h)}
            \leq \frac{4}{\ln(4/3)}
            \leq 14.
        \end{equation*}
    \end{corollary}
\end{samepage}

The proof of Corollary~\ref{thm:approx} is presented in
Section~\ref{sec:proofs_approx}. We point out that choosing the duty
cycle $\delta^\star$ as a function of $N$, $g$, and $h$, is not
necessary to obtain the additive approximation result in
Corollary~\ref{thm:approx}. In fact, using only simple
amplify-and-forward achieves the same additive approximation guarantee,
i.e.,
\begin{equation*}
    C(N,g,h)-R_1(N,g,h)
    \leq 1.8 \text{ bits}
\end{equation*}
for all $N\geq 2$, $g,h > 0$. However, the same is not true if we are
also interested in multiplicative approximation guarantees (at least in
the low-rate regime). To achieve a constant additive approximation as
well as constant multiplicative approximation, the duty cycle
$\delta^\star$ is required to vary as a function of $N$, $g$, and $h$,
and therefore bursty amplify-and-forward is required.

From Theorems~\ref{thm:achievability} and \ref{thm:converse}, the
capacity of the symmetric diamond network has three distinct regimes,
depending on whether $g \leq h$, $h < g < N^2h$, or $g\geq N^2h$.  In
the first regime ($g\leq h$), the channel gain to the relays is weak
compared to the channel gain to the destination, and the achievable rate
is constrained by the broadcast part of the diamond network. The
capacity in this regime is given approximately by
\begin{equation*}
    C(N,g,h) \approx \tfrac{1}{2}\log(1+Ng),
\end{equation*}
where the approximation is in the sense of Corollary~\ref{thm:approx},
namely up to a multiplicative gap of factor $14$ in the low-rate regime
($g \ll N^{-1}$) and up to an additive gap of $1.8$ bits in the
high-rate regime ($g \gg N^{-1}$).  This is the capacity of a
single-input multiple-output channel with unit power constraint, one
transmit antenna, $N$ receive antennas, and channel gain $\sqrt{g}$
between each of them.  Thus, the broadcast cut in
Fig.~\ref{fig:diamond_cuts1} in Section~\ref{sec:intro} is approximately
tight in this regime.

In the third regime ($g\geq N^2h$), the channel gain to the relays is
strong compared to the channel gain to the destination, and the
achievable rate is now constrained by the multiple-access part of the
channel. The capacity in the third regime is given approximately by
\begin{equation*}
    C(N,g,h) \approx \tfrac{1}{2}\log(1+N^2h).
\end{equation*}
This is the capacity of a multiple-input single-output channel with unit
per-antenna power constraint, $N$ transmit antennas, one receive
antenna, and channel gain $\sqrt{h}$ between each of them. Thus, the
multiple-access cut in Fig.~\ref{fig:diamond_cuts1} is approximately
tight in this regime.  Observe that to achieve this rate the signals
sent by the relays must be highly correlated and add up coherently at
the destination.

The most interesting regime is the second one ($h < g < N^2h$). If 
$\max\{g,Nh\}\geq 1$, then the capacity is given approximately by 
\begin{equation*}
    C(N,g,h) \approx \tfrac{1}{2}\log\big(1+N\min\{g,Nh\}\big),
\end{equation*}
and again either the broadcast cut or the multiple-access cut are tight.
If $\max\{g,Nh\} < 1$ the situation is more complicated. If
$N\sqrt{gh}\geq 1$, then the capacity of the diamond network is
approximately
\begin{equation*}
    C(N,g,h) \approx \tfrac{1}{2}\log(1+N^2gh), 
\end{equation*}
and, if $N\sqrt{gh}< 1$,
\begin{equation*}
    C(N,g,h) \approx \tfrac{1}{2}\log\big(1+N{\textstyle\sqrt{gh}}\big).
\end{equation*}
In both cases, the capacity depends on the product of $g$ and $h$, and
not merely on the minimum of $g$ and $Nh$. Hence, neither the broadcast
cut nor the multiple-access cut are tight in this case. In fact, these
bounds can be arbitrarily bad, both in terms of additive gap as well as
multiplicative gap, as the next two examples illustrate.

For the additive gap, consider $g=N^{-5/8}$ and $h=N^{-9/8}$. Then
$\max\{g,Nh\}= N^{-1/8} < 1$, $g=N^{1/2}h\in(h,N^2h)$, and $N\sqrt{gh} =
N^{1/8} \geq 1$, so that
\begin{align*}
    C(N,g,h) 
    & \approx \tfrac{1}{2}\log(1+N^2gh) \\
    & = \tfrac{1}{2}\log(1+N^{1/4}). 
\end{align*}
On the other hand, the minimum of the broadcast and multiple-access
cuts yields
\begin{equation*}
    \tfrac{1}{2}\log\big(1+N\min\{g,Nh\}\big)
    = \tfrac{1}{2}\log(1+N^{3/8}),
\end{equation*}
resulting in an additive gap of order $\Theta(\log(N))$ bits, which is
unbounded as the number of relays $N\to\infty$.

For the multiplicative gap, consider $g=N^{-2}$ and $h=N^{-3}$.
Then $\max\{g,Nh\}= N^{-2} < 1$, $g=Nh\in(h,N^2h)$, and
$N\sqrt{gh} = N^{-3/2} < 1$, so that
\begin{align*}
    C(N,g,h) 
    & \approx \tfrac{1}{2}\log\big(1+N{\textstyle\sqrt{gh}}\big) \\
    & = \tfrac{1}{2}\log(1+N^{-3/2}) \\
    & \approx \tfrac{1}{2}\log(e)N^{-3/2}.
\end{align*}
On the other hand, the minimum of the broadcast and multiple-access
cuts yields
\begin{align*}
    \tfrac{1}{2}\log\big(1+N\min\{g,Nh\}\big)
    & = \tfrac{1}{2}\log(1+N^{-1}) \\ 
    & \approx \tfrac{1}{2}\log(e)N^{-1},
\end{align*}
resulting in a multiplicative gap of order $\Theta(\sqrt{N})$,
which is again unbounded as the number of relays $N\to\infty$.

\begin{figure}[htbp]
    \begin{center}
        \scalebox{1}{\input{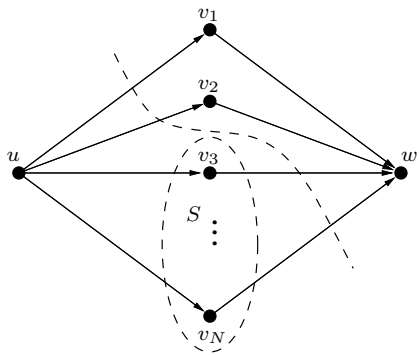}}
    \end{center}

    \caption{A general cut in the diamond network. Here
    $S\subset\{1,\ldots,N\}$, and the cut separates $u\cup\{v_n\}_{n\in S}$
    from $w\cup\{v_n\}_{n\in S^c}$.}

    \label{fig:diamond_cuts2}
\end{figure}

In the second regime, we thus need to take cuts other than the broadcast
and multiple-access ones into account. The need for this can be
understood as follows. Consider a general cut separating the source node
$u$ from the destination node $w$ in the diamond network as shown in
Fig.~\ref{fig:diamond_cuts2}. 
Formally, let $S\subset \{1,\ldots, N\}$, and consider the cut from
$u\cup\{v_n\}_{n\in S}$ to $w\cup\{v_n\}_{n\in S^c}$. Assume the
signals $(X_n)_{n=1}^N$ sent from the relays to the destination are
highly correlated. This results in the signal summing up coherently at
the receiver, increasing the rate across the cut. At the same time, if
the signals sent from the relays are highly correlated, then the signals
$(X_n)_{n\in S^c}$ available at the relays on the other side of the
cut can be used to estimate the signal received at the destination node.
This decreases the rate across the cut. Thus, for general cuts, there is
a tradeoff between the gain from coherent reception and the loss from
prediction that come with increased signal correlation. This tradeoff is
absent if we only consider the broadcast and multiple-access cuts. It is
precisely this tradeoff that determines the behavior of the capacity of
the diamond network in the second regime.

We point out that a (partial) decode-and-forward strategy is
not sufficient to provide a uniform capacity approximation as in
Corollary~\ref{thm:approx}. Indeed, due to symmetry, \emph{all} relays
would be able to decode the source in any such strategy, which implies
that decode-and-forward and partial decode-and-forward coincide in this
case. The rate achievable with decode-and-forward is given by
\begin{equation*}
    \tfrac{1}{2}\log\big(1+\min\{g, N^2h\}\big).
\end{equation*}
Comparing this with Corollary~\ref{thm:approx}, we see that (partial)
decode-and-forward has an additive gap of at least $\Omega(\log(N))$
bits and a multiplicative gap of at least a factor $\Omega(N)$ to
capacity. Similarly, compress-and-forward does not achieve constant (in
the network size $N$) additive or multiplicative gaps to capacity, since
it does not capture the gain from coherent signal addition at the
destination. Finally, as was pointed out earlier, the traditional
amplify-and-forward strategy does not yield a constant factor
approximation of capacity. In fact, in can be shown that simple
amplify-and-forward results in unbounded multiplicative gap even for
$N=2$. Therefore the bursty amplify-and-forward scheme introduced in
\cite{schein01} and advocated in this work has the nice property of
being uniformly approximately optimal in both the additive sense and the
multiplicative sense, as well as being a simple modification of the
traditional amplify-and-forward scheme.

\subsection{Asymmetric Diamond Networks}
\label{sec:main_asymmetric}

In the last section, we have considered \emph{symmetric} diamond
networks, in which the channel gain from the source $u$ to the relay
$v_n$ is $\sqrt{g}$ and the channel gain from $v_n$ to the destination
$w$ is $\sqrt{h}$ for all $n$. In this section, we show how some of the
results can be extended to \emph{asymmetric} diamond networks, in which
the channel gains $(g_n)_{n=1}^N$ and $(h_n)_{n=1}^N$ are allowed to
take arbitrary values.

For this asymmetric setting, it was shown in~\cite{avestimehr09} that
(bursty) amplify-and-forward does \emph{not} achieve a constant (as a
function of $g_n$ and $h_n$) additive-gap approximation even when $N=2$.
However, we show here that bursty amplify-and-forward is approximately
optimal in the sense of multiplicative approximation for any $N$,
$(g_n)_{n=1}^N$, and $(h_n)_{n=1}^N$. More precisely, we show that the
rate achieved by bursty amplify-and-forward combined with optimal relay
selection is at most a factor $O(\log^4(N))$ from capacity uniformly in
$(g_n)_{n=1}^N$ and $(h_n)_{n=1}^N$. While not constant in $N$, compared
to the best previously known multiplicative approximation of a factor
$2N(N+1)$, this is a significant improvement.  Hence, at least in the
low-rate regime, bursty amplify-and-forward is also a good communication
scheme for asymmetric diamond networks.

\begin{theorem}
    \label{thm:asymmetric}
    There exists a universal constant $K < \infty$ such that for every
    diamond network with $N\geq 2$ relays and channel gains 
    $(g_n)_{n=1}^N, (h_n)_{n=1}^N> 0$, 
    \begin{equation*}
        \frac{C\big(N,(g_n),(h_n)\big)}
        {\sup_{\delta\in(0,1]}R_\delta\big(N,(g_n),(h_n)\big)}
        \leq K\log^4(N).
    \end{equation*}
\end{theorem}

The proof of Theorem~\ref{thm:asymmetric} is presented in
Section~\ref{sec:proofs_asymmetric}. At a high level, achievability is
proved as follows. Group the relays into classes such that all relays in
the same class have approximately the same channel gains. Choose one
such class, and set the constants $\alpha_n=0$ for all relays not in
this class (effectively disabling those relays). This relay-selection
step reduces the original asymmetric network to a (almost) symmetric
subnetwork. Theorem~\ref{thm:achievability} can now be applied to this
symmetric subnetwork to obtain a lower bound on the rate achievable with
bursty amplify-and-forward. By maximizing over all possible classes, and
hence all possible symmetric subnetworks, we get the largest rate
achievable in this manner. The corresponding upper bound shows that this
approach of relay selection combined with bursty amplify-and-forward is
approximately optimal.

\section{Proofs}
\label{sec:proofs}

Sections~\ref{sec:proofs_achievability}, \ref{sec:proofs_converse}, and
\ref{sec:proofs_approx} contain the proofs of
Theorem~\ref{thm:achievability} (achievability),
Theorem~\ref{thm:converse} (upper bound), and Corollary~\ref{thm:approx}
(approximation) for symmetric diamond networks.
Section~\ref{sec:proofs_asymmetric} contains the proof of
Theorem~\ref{thm:asymmetric} for general asymmetric diamond networks.

\subsection{Proof of Theorem~\ref{thm:achievability} (Achievability for
Symmetric Networks)}
\label{sec:proofs_achievability}

We start with a lemma computing the rate achievable by
amplify-and-forward.

\begin{lemma}
    \label{thm:r}
    For every symmetric diamond network with $N\geq 2$ relays and channel gains $g,h> 0$, 
    amplify-and-forward achieves 
    \begin{equation*}
        R_1(N,g,h) = \tfrac{1}{2}\log\Big(1+\frac{N^2gh}{1+g+Nh}\Big).
    \end{equation*}
\end{lemma}
\begin{IEEEproof}
    Recall that with amplify-and-forward relay $v_n$ transmits
    \begin{equation*}
        X_n[t] 
        = \alpha Y_n[t]
        = \alpha \sqrt{g}X[t] + \alpha Z_n[t]
    \end{equation*} 
    at time $t$, with constant $\alpha\geq 0$ chosen to satisfy the
    average unit power constraint. The received signal at the
    destination node $w$ is 
    \begin{equation*}
        Y[t] 
        = \alpha N {\textstyle \sqrt{gh}} X[t] 
        + \alpha{\textstyle \sqrt{h\vphantom{g}}}\sum_{n=1}^N Z_n[t] + Z[t].
    \end{equation*}
    Observe that this describes a memoryless point-to-point
    channel with channel gain $\alpha N \sqrt{gh}$ and additive Gaussian
    noise of variance $1+\alpha^2 N h$. $R_1(N,g,h)$ is the capacity of
    this point-to-point channel, optimized over the value of $\alpha$.

    For any value of $\alpha\geq 0$, the optimal distribution of the
    input $X$ for this point-to-point channel is Gaussian with zero mean and
    variance one.  The signal sent by the relays has power
    \begin{equation*}
        \E(X_n^2) = \alpha^2(1+g),
    \end{equation*}
    and hence for
    \begin{equation*}
        \alpha^2 \in \big[0,1/(1+g)\big]
    \end{equation*}
    the average unit power constraints at the relays are satisfied.
    This implies that amplify-and-forward achieves a rate of
    \begin{align*}
        R_1(N,g,h)
        & = \max_{\alpha^2\in[0,1/(1+g)]}
        \tfrac{1}{2}\log\Big(1+\frac{\alpha^2 N^2gh}{1+\alpha^2Nh}\Big) \\
        & = \tfrac{1}{2}\log\Big(1+\frac{N^2gh}{1+g+Nh}\Big).
    \end{align*}
\end{IEEEproof}

The next lemma describes the rate achievable with bursty
amplify-and-forward.

\begin{lemma}
    \label{thm:rdelta}
    For every symmetric diamond network with $N\geq 2$ relays and
    channel gains $g,h > 0$, bursty amplify-and-forward with duty cycle
    $\delta\in(0,1]$ achieves
    \begin{equation*}
        R_\delta(N,g,h)
        = \tfrac{1}{2}\delta\log\Big(1+\frac{N^2gh/\delta^2}{1+g/\delta+Nh/\delta}\Big).
    \end{equation*}
\end{lemma}
\begin{IEEEproof}
    During the $\delta$ fraction of time we communicate, we are dealing
    with an equivalent channel with gains
    $\sqrt{\smash[b]{g}\smash{/}\delta}$,
    $\sqrt{h\smash{/}\delta}$ and with unit power constraints. The result now follows
    from Lemma~\ref{thm:r} by taking into account that we only
    communicate a fraction $\delta$ of time.
\end{IEEEproof}

Note that Lemmas~\ref{thm:r} and \ref{thm:rdelta} coincide for
$\delta=1$, as expected.  We now proceed to the proof of
Theorem~\ref{thm:achievability}. To simplify notation, set
\begin{equation*}
    R_\delta \defeq R_\delta(N,g,h)
\end{equation*}
for $\delta\in (0,1]$. 

We consider the cases $\max\{g,Nh\}\geq 1$ and $\max\{g,Nh\} < 1$
separately. Assume first $\max\{g,Nh\} \geq 1$. Here we set
$\delta = 1$, i.e., we use simple amplify-and-forward. By
Lemma~\ref{thm:r}
\begin{align*}
    R_1 
    & = \tfrac{1}{2}\log\Big(1+\frac{N^2gh}{1+g+Nh}\Big) \\
    & = \tfrac{1}{2}\log\bigg(1+
    \frac{N\min\{g,Nh\}\max\{g,Nh\}}{1+\min\{g,Nh\}+\max\{g,Nh\}}\bigg) \\
    & \geq \tfrac{1}{2}\log\bigg(1+
    \frac{N\min\{g,Nh\}\max\{g,Nh\}}{3\max\{g,Nh\}}\bigg) \\
    & = \tfrac{1}{2}\log\big(1+\tfrac{1}{3}N\min\{g,Nh\}\big),
\end{align*}
where we have used that $1 \leq \max\{g,Nh\}$ to obtain the inequality.

Assume in the following that $\max\{g,Nh\} < 1$. We consider the cases
$g\leq h$, $g\in(h,N^2h)$, and $g\geq N^2h$ separately.  Consider first
$g \leq h$. Bursty amplify-and-forward with duty cycle $\delta=Ng \leq
Nh \leq 1$ achieves by Lemma~\ref{thm:rdelta}
\begin{align*}
    R _{\delta} 
    & = \tfrac{1}{2}Ng
    \log\Big(1+\frac{N^2gh/(N^2g^2)}{1+g/(Ng)+Nh/(Ng)}\Big) \\
    & = \tfrac{1}{2}Ng\log\Big(1+\frac{h}{g+g/N+h}\Big) \\
    & \stackrel{(a)}{\geq} \tfrac{1}{2}Ng\log\Big(1+\frac{h}{h+h/N+h}\Big) \\
    & \geq \tfrac{1}{2}Ng \log(4/3) \\
    & \geq \tfrac{1}{2}\ln(4/3)\log(1+Ng),
\end{align*}
where in $(a)$ we used $g \leq h$.

Consider then $g\in(h,N^2h)$. If $N\sqrt{gh} \geq 1$, then simple
amplify-and-forward achieves by Lemma~\ref{thm:r}
\begin{align*}
    R_1
    & = \tfrac{1}{2}\log\Big(1+\frac{N^2gh}{1+g+Nh}\Big) \\
    & \geq \tfrac{1}{2}\log\big(1+\tfrac{1}{3}N^2gh\big),
\end{align*}
where we have used that $1+g+Nh\leq 3$, which follows from $\max\{g,Nh\}\leq 1$. 

Still assuming $g\in(h,N^2h)$, if $N\sqrt{gh} < 1$,\footnote{Note that
$g\in(h,N^2h)$ and $N\sqrt{\smash{g}h}< 1$ imply $\max\{g,Nh\}< 1$.}
then bursty amplify-and-forward with duty cycle $\delta=N\sqrt{gh}\leq
1$ achieves by Lemma~\ref{thm:rdelta}
\begin{align*}
    R_\delta
    & = \tfrac{1}{2}N{\textstyle\sqrt{gh}}\log\Big(1+\frac{N^2gh/(N^2gh)}{\textstyle 1+g/(N\sqrt{gh})+Nh/(N\sqrt{gh})}\Big) \\
    & = \tfrac{1}{2}N{\textstyle\sqrt{gh}}\log\Big(1
    +\frac{1}{\textstyle 1+\sqrt{g\vphantom{h}}/(N\sqrt{h\vphantom{g}})+\sqrt{h\vphantom{g}}/\sqrt{g\vphantom{h}}}\Big) \\
    & \stackrel{(b)}{\geq} \tfrac{1}{2}N{\textstyle\sqrt{gh}}\log\Big(1
    +\frac{1}{\textstyle 1+\sqrt{N^2h}/(N\sqrt{h})+\sqrt{h}/\sqrt{h}}\Big) \\
    & = \tfrac{1}{2}N{\textstyle\sqrt{gh}}\log(4/3) \\
    & \geq \tfrac{1}{2}\ln(4/3)\log\big(1+N{\textstyle\sqrt{gh}}\big),
\end{align*}
where in $(b)$ we have used that $g\leq N^2h$ and $g\geq h$.

Consider finally $g \geq N^2h$. Bursty amplify-and-forward with duty
cycle $\delta=N^2h\leq g \leq 1$ achieves by Lemma~\ref{thm:rdelta}
\begin{align*}
    R_\delta
    & = \tfrac{1}{2}N^2h\log\Big(1+\frac{N^2gh/(N^4h^2)}{1+g/(N^2h)+Nh/(N^2h)}\Big) \\
    & = \tfrac{1}{2}N^2h\log\Big(1+\frac{g/(N^2h)}{1+g/(N^2h)+1/N}\Big) \\
    & \stackrel{(c)}{\geq} \tfrac{1}{2}N^2h\log(4/3) \\
    & \geq \tfrac{1}{2}\ln(4/3)\log(1+N^2h), 
\end{align*}
where in $(c)$ we have used that $1/N\leq 1 \leq g/(N^2h)$.
\hfill\IEEEQED

\subsection{Proof of Theorem~\ref{thm:converse} (Upper Bound for
Symmetric Networks)}
\label{sec:proofs_converse}

In this section, we derive an upper bound on the capacity of the Gaussian
diamond network. The standard way to find such bounds is to start with
the cut-set bound and then to simplify it further to obtain a closed-form
expression. The derivation here starts with the cut-set bound as well,
but differs in several key aspects from the standard approach, which we
now highlight.

Let
\begin{equation*}
    [N] \defeq \{1,2,\ldots, N\},
\end{equation*}
and for a subset $S\subset [N]$, define
\begin{equation*}
    S^c \defeq [N]\setminus S.
\end{equation*}
By the cut-set bound \cite[Theorem 14.10.1]{cover91}, 
\begin{equation*}
    C(N,g,h) 
    \leq \sup_{X,X_{[N]}} \min_{S\subset [N]} 
    I\big(X, X_S; Y, Y_{S^c}\bigm\vert X_{S^c}\big),
\end{equation*}
where the maximization is over random variables $X, X_{[N]}$ satisfying
the power constraints $\E(X^2) \leq 1$, $\E(X_n^2)\leq 1$, and where
$X_{\tilde{S}} \defeq (X_n)_{n\in \tilde{S}}$ for any subset
$\tilde{S}\subset [N]$ (see Fig.~\ref{fig:diamond_cuts2} in
Section~\ref{sec:main_symmetric}).
A short calculation (done in \eqref{eq:converse2} below) reveals that
\begin{equation}
    \label{eq:cutset}
    \sup_{X,X_{[N]}} \min_{S\subset [N]}
    I\big(X, X_S; Y, Y_{S^c}\bigm\vert X_{S^c}\big)
    \leq \sup_{X,X_{[N]}} \min_{S\subset [N]}
    \Big(I\big(X; Y_{S^c}\big)+I\big(X_S; Y\bigm\vert X_{S^c}\big)\Big).
\end{equation}
In the right-hand side of \eqref{eq:cutset}, the first mutual
information corresponds to the rate between the source nodes and the
relays, and the second mutual information corresponds to the rate
between the relays and the destination node. 

One approach is to simplify this expression further through a sequence
of two steps. The first step is to upper bound
\begin{align*}
    I\big(X_S; Y\bigm\vert X_{S^c}\big)
    & = \mc{H}\big(Y \bigm\vert X_{S^c}\big)-\mc{H}(Z) \\
    & \leq \mc{H}\Big({\textstyle\sqrt{h}\sum_{n\in S}}X_n+Z\Big)-\mc{H}(Z) \\
    & = I\Big(X_S; {\textstyle\sqrt{h}\sum_{n\in S}}X_n+Z \Big),
\end{align*}
where, in order to avoid confusion with the channel gain $h$, we denote
the differential entropy by the non-standard symbol $\mc{H}$. 
This first step thus removes the conditioning on the signals
$X_{S^c}$ available at the destination side of the cut. The second step
is to interchange the order of maximization and minimization. This
yields
\begin{align}
    \label{eq:weak1}
    C(N,g,h) 
    & \leq \min_{S\subset [N]} \sup_{X,X_{[N]}} 
    \Big(I\big(X; Y_{S^c}\big)+I\Big(X_S; {\textstyle\sqrt{h}\sum_{n\in S}}X_n +Z\Big)\Big) \nonumber\\
    & = \min_{n\in \{0,\ldots,N\}} \big(
    \tfrac{1}{2}\log(1+(N-n)g) + \tfrac{1}{2}\log(1+n^2h)
    \big).
\end{align}
This can be further upper bounded by considering only $n=0$ or $n=N$,
resulting in the minimum of the broadcast and multiple-access cut
\begin{equation}
    \label{eq:weak2}
    C(N,g,h)
    \leq \min\big\{ \tfrac{1}{2}\log(1+Ng), \tfrac{1}{2}\log(1+N^2h) \big\}.
\end{equation}
Neither of the upper bounds \eqref{eq:weak1} and \eqref{eq:weak2} are
tight enough to obtain a constant gap approximation of the capacity
(this can be seen from the two examples presented after
Corollary~\ref{thm:approx}).

In this paper, we also start the derivation of the upper bound from the
cut-set bound \eqref{eq:cutset}, but we avoid taking the two simplifying
steps mentioned in the last paragraph. Instead, we first show, using the
symmetry in the problem, that the correlation between any two signals
$X_{n}$ and $X_{{\tilde{n}}}$ with $n\neq \tilde{n}$ can be assumed
to be equal without loss of optimality.  Using the resulting simple form
of the covariance matrix allows us then to evaluate the term $I\big(X_S;
Y\vert X_{S^c}\big)$ directly. This enables us to keep the conditioning
on $X_{S^c}$, which yields a significantly tighter upper bound on
capacity. The resulting upper bound is summarized in the following
lemma.

\begin{lemma}
    \label{thm:cutset}
    For every symmetric diamond network with $N\geq 2$ relays and
    channel gains $g, h > 0$, capacity is upper bounded as
    \begin{align*}
        C(N, & g,h) \\
        & \leq \sup_{\rho\in[0,1)} \min_{n\in \{0,\ldots, N\}}
        \biggl(\tfrac{1}{2}\log(1+(N-n)g) 
        + \tfrac{1}{2}\log
        \Bigl(1+ n\Bigl(1+(n-1)\rho-\frac{n(N-n)\rho^2}{1+(N-n-1)\rho}\Bigr)h
        \Bigr)
        \biggr).
    \end{align*}
\end{lemma}
The variable $\rho$ appearing in the lemma can be interpreted as the
correlation between the random variables $X_{[N]}$ as mentioned in the
preceding discussion. Note that it is not clear a priori that this
correlation $\rho$ can be restricted to be nonnegative. This restriction
is part of the assertion of the lemma. We also point out that it is important that
$\rho=1$ is excluded from the supremum in Lemma~\ref{thm:cutset}; the
result is not true without this restriction.

It will be convenient in the following to work with a weaker version of
Lemma~\ref{thm:cutset}. Note that, for $\rho\in[0,1)$,
\begin{align*}
    1+(n-1)\rho - \frac{n(N-n)\rho^2}{1+(N-n-1)\rho} 
    & \leq 1+n\rho-\frac{n(N-n)\rho^2}{1+(N-n)\rho} \\
    & = \Big(\frac{N}{N-n}\Big)
    \Big(\frac{\frac{N-n}{N}+(N-n)\rho}{1+(N-n)\rho}\Big) \\
    & \leq \frac{N}{N-n}.
\end{align*}
Hence
\begin{equation}
    \label{eq:cutset_simple}
    C(N,g,h) 
    \leq \min_{n\in \{0,\ldots, N\}} 
    \bigg(\tfrac{1}{2}\log(1+(N-n)g) 
    + \tfrac{1}{2}\log\Big(1+\frac{N^2}{N-n}h\Big)\bigg).
\end{equation}

The upper bound \eqref{eq:cutset_simple} derived from
Lemma~\ref{thm:cutset} can be compared to the simpler bound
\eqref{eq:weak1}. If $n=KN$ for some constant $K\in(0,1)$, then the
factor multiplying the channel gain $h$ in \eqref{eq:weak1} is of order
$\Theta(N^2)$. On the other hand, the same factor in
\eqref{eq:cutset_simple} is of order $\Theta(N)$. Thus, the bound
\eqref{eq:cutset_simple} can be considerably tighter than the simpler
bound \eqref{eq:weak1}.

\begin{IEEEproof}[Proof of Lemma~\ref{thm:cutset}]
    By the cut-set bound \cite[Theorem 14.10.1]{cover91}, 
    \begin{equation}
        \label{eq:converse1}
        C \defeq C(N,g,h) 
        \leq \sup_{X,X_{[N]}} \min_{S\subset [N]} 
        I\big(X, X_S; Y, Y_{S^c}\bigm\vert X_{S^c}\big),
    \end{equation}
    where, as before, the maximization is over random variables $X,
    X_{[N]}$ satisfying the power constraints $\E(X^2) \leq 1$,
    $\E(X_n^2)\leq 1$. We evaluate \eqref{eq:converse1} in two steps.
    First, we argue that the maximization over $X, X_{[N]}$ can be
    restricted to jointly Gaussian random variables such that each
    $\E(X_n^2) = 1$ and $\E(X_n X_{\tilde{n}})=\rho$ for
    $n\neq\tilde{n}$ and some $\rho\in[-1/(N-1),1]$. This simplifies the
    maximization to be over just the parameter $\rho$ instead of
    $N$-dimensional distributions.  Second, using the resulting simple
    form of the input distributions, we analytically evaluate the mutual
    information in \eqref{eq:converse1} to obtain the stated bound.

    We start by simplifying the mutual information in
    \eqref{eq:converse1} for a fixed cut $S\subset [N]$. We have
    \begin{align}
        \label{eq:converse2}
        I\big( X, X_S; & Y, Y_{S^c}\bigm\vert X_{S^c}\big) \nonumber\\
        & = \mc{H}\big(Y, Y_{S^c}\bigm\vert X_{S^c}\big)
        - \mc{H}\big(Y, Y_{S^c}\bigm\vert X, X_{[N]}\big) \nonumber\\ 
        & = \mc{H}\big(Y_{S^c}\bigm\vert X_{S^c}\big)
        + \mc{H}\big(Y\bigm\vert Y_{S^c}, X_{S^c}\big)
        - \mc{H}\big(Y_{S^c}\bigm\vert X, X_{[N]}\big)
        - \mc{H}\big(Y\bigm\vert Y_{S^c}, X, X_{[N]}\big) \nonumber\\
        & \leq \mc{H}\big(Y_{S^c}\big)
        + \mc{H}\big(Y\bigm\vert X_{S^c}\big)
        - \mc{H}\big(Y_{S^c}\bigm\vert X\big)
        - \mc{H}\big(Y\bigm\vert X_{[N]}\big) \nonumber\\
        & = I\big(X; Y_{S^c}\big)+I\big(X_S; Y\bigm\vert X_{S^c}\big),
    \end{align}
    where we have used that
    \begin{align*}
        \mc{H}\big(Y_{S^c}\bigm\vert X, X_{[N]}\big)
        = \mc{H}\big(Z_{S^c} \big) 
        = \mc{H}\big(Y_{S^c}\bigm\vert X\big),
    \end{align*}
    and that
    \begin{align*}
        \mc{H}\big(Y\bigm\vert Y_{S^c}, X, X_{[N]}\big) 
        = \mc{H}(Z)
        = \mc{H}\big(Y\bigm\vert X_{[N]}\big).
    \end{align*}
    Combining \eqref{eq:converse1} and \eqref{eq:converse2} yields
    \begin{equation}
        \label{eq:converse3}
        C 
        \leq \sup_{X,X_{[N]}} \min_{S\subset [N]} 
        \Big(I\big(X; Y_{S^c}\big)+I\big(X_S; Y\bigm\vert X_{S^c}\big)\Big).
    \end{equation}

    For the first term in \eqref{eq:converse3},
    \begin{equation}
        \label{eq:converse4}
        I\big(X; Y_{S^c}\big)
        \leq \tfrac{1}{2}\log(1+\card{S^c}g),
    \end{equation}
    since the channel from $X$ to $Y_{S^c}$ is a Gaussian
    single-input multiple-output channel with channel gains $\sqrt{g}$.
    For the second term in \eqref{eq:converse3},
    \begin{align}
        \label{eq:converse5}
        I\big(X_S; Y\bigm\vert X_{S^c}\big)
        & = \mc{H}\big(Y \bigm\vert X_{S^c}\big)
        - \mc{H}\big(Y\bigm\vert X_{[N]}\big) \nonumber\\
        & = \mc{H}\Big({\textstyle\sqrt{h}\sum_{n\in S}}\big(X_n - \beta_n(X_{S^c})\big)+Z 
        \Bigm\vert X_{S^c}\Big)- \mc{H}(Z) \nonumber\\
        & \leq \mc{H}\Big({\textstyle\sqrt{h}\sum_{n\in S}} \big( X_n - \beta_n(X_{S^c})\big)+Z\Big)
        - \mc{H}(Z),
    \end{align}
    for any choice of functions $\beta_n(X_{S^c})$ for $n\in S$. In
    particular, let $\beta_n(X_{S^c})$ be the minimum mean-square error
    estimator for $X_n$ based on $X_{S^c}$. 
    
    Let $X_{[N]}$ have covariance matrix $\bm{Q}$.  Then, by
    \cite[Theorem 1.2.11]{muirhead82}, $(X_n - \beta_n(X_{S^c}))_{n\in
    S}$ has covariance matrix 
    \begin{equation}
        \label{eq:moore-penrose}
        \bm{Q}_{S\vert S^c}
        \defeq \bm{Q}_{S,S}-\bm{Q}_{S, S^c}\bm{Q}_{S^c,S^c}^-\bm{Q}_{S^c, S},
    \end{equation}
    where, for any subsets $S_1, S_2\subset [N]$, $\bm{Q}_{S_1,S_2}$ is
    the submatrix of $\bm{Q}$ induced by the rows $S_1$ and columns
    $S_2$, and where $\bm{Q}_{S^c,S^c}^-$ is the \emph{Moore-Penrose
    generalized inverse} of the matrix $\bm{Q}_{S^c,S^c}$. The matrix
    $\bm{Q}_{S\vert S^c}$ is called the \emph{generalized Schur
    complement} of $\bm{Q}_{S^c, S^c}$ in $\bm{Q}$. Note that if
    $\bm{Q}_{S^c,S^c}$ is invertible, then $\bm{Q}_{S^c,S^c}^- =
    \bm{Q}_{S^c,S^c}^{-1}$ and the generalized Schur complement reduces
    to the standard Schur complement. 

    Before proceeding, we need to introduce some notation.  Denote by
    $\bm{I}_a$ the $a\times a$ identity matrix, and by $\bm{1}_{a,b}$
    the $a\times b$ matrix of ones. To simplify notation, we will
    write $\bm{1}$ for the column vector $\bm{1}_{a,1}$, whenever the
    dimension is clear from the context. With these definitions,
    \begin{equation}
        \label{eq:converse6}
        \mc{H}\Big({\textstyle\sqrt{h}\sum_{n\in S}} \big( X_n - \beta_n(X_{S^c})\big)+Z\Big)
        - \mc{H}(Z)
        \leq \tfrac{1}{2}\log\big(1
        +h\bm{1}^T\bm{Q}_{S\vert S^c}\bm{1}\big).
    \end{equation}

    Substituting \eqref{eq:converse4}, \eqref{eq:converse5}, and
    \eqref{eq:converse6} into \eqref{eq:converse3} yields
    \begin{align*}
        C \leq \sup_{\substack{\bm{Q}\geq 0:\\ q_{n,n}\leq 1 \forall n\in [N]}} 
        \min_{S\subset [N]} 
        \Big(\tfrac{1}{2}\log(1+\card{S^c}g)
        + \tfrac{1}{2}\log\big(1
        +h\bm{1}^T\bm{Q}_{S\vert S^c}\bm{1}\big)
        \Big),
    \end{align*}
    where $\bm{Q}\geq 0$ denotes that $\bm{Q}$ is a positive semi-definite
    matrix. We have thus simplified the maximization over input distributions to
    a maximization over covariance matrices. The next step is to show
    that the covariance matrix $\bm{Q}$ can be restricted without loss
    of optimality to have the form 
    \begin{equation*}
        \rho \bm{1}_{N,N} + (1-\rho) \bm{I}_{N},
    \end{equation*}
    and hence the maximization over covariance matrices can be further
    simplified to a maximization over just the scalar correlation parameter
    $\rho$.\footnote{Upon completion of this work, we realized that a
    somewhat similar argument as in this step was used in 
    \cite[Section~III]{thomas87} for the Gaussian multiple-access
    channel with feedback.}
 
    For convenience of notation, define
    \begin{equation*}
        \psi_S(\bm{Q})
        \defeq \tfrac{1}{2}\log\big(1+h\bm{1}^T\bm{Q}_{S\vert S^c}\bm{1}\big)
    \end{equation*}
    and
    \begin{equation*}
        \psi(\bm{Q})
        \defeq \min_{S\subset [N]} \Big(\tfrac{1}{2}\log(1+\card{S^c}g) + \psi_S(\bm{Q}) \Big),
    \end{equation*}
    so that
    \begin{equation}
        \label{eq:converse7}
        C \leq \sup_{\substack{\bm{Q}\geq 0:\\ q_{n,n}\leq 1 \forall n\in [N]}}
        \psi(\bm{Q}).
    \end{equation}
    Consider a covariance matrix $\bm{Q}\geq 0$, and let $\bm{P}$ be any
    permutation matrix on $[N]$. Note that $\bm{P}^T\bm{Q}\bm{P}\geq 0$.
    Moreover, by symmetry,\footnote{Note that the minimization over
    $S\subset [N]$ is crucial for this fact to hold. Indeed,
    $\psi_S(\bm{Q})\neq\psi_S\big(\bm{P}^T\bm{Q}\bm{P}\big)$ in
    general.}
    \begin{align*}
        \psi(\bm{Q})
        & = \min_{S\subset [N]} \big(\tfrac{1}{2}\log(1+\card{S^c}g) + \psi_S(\bm{Q}) \big) \\
        & = \min_{S\subset [N]} \big(\tfrac{1}{2}\log(1+\card{S^c}g) 
        + \psi_S\big(\bm{P}^T\bm{Q}\bm{P}\big) \big) \\
        & = \psi\big(\bm{P}^T\bm{Q}\bm{P}\big),
    \end{align*}
    and thus $\psi(\cdot)$ is invariant under permutation.

    Now, the generalized Schur complement is matrix-concave over the set
    of positive semi-definite matrices \cite[Theorem 3.1]{li00} (see
    also \cite[p.  469]{marshall79} for the corresponding result for
    positive definite matrices). More precisely, if $\bm{Q} = \lambda_1
    \bm{Q}^1+ \lambda_2\bm{Q}^2$ with $\lambda_1\in[0,1]$,
    $\lambda_2=1-\lambda_1$, then
    \begin{equation*}
        \bm{Q}_{S\vert S^c}
        \geq \lambda_1 \bm{Q}_{S\vert S^c}^1+ \lambda_2\bm{Q}_{S\vert
        S^c}^2,
    \end{equation*}
    i.e., 
    \begin{equation*}
        \bm{Q}_{S\vert S^c}
        - \big(\lambda_1 \bm{Q}_{S\vert S^c}^1+ \lambda_2\bm{Q}_{S\vert
        S^c}^2\big)
    \end{equation*}
    is a positive semi-definite matrix. Therefore, 
    \begin{equation*}
        \bm{1}^T\big(\bm{Q}_{S\vert S^c}
        - \big(\lambda_1 \bm{Q}_{S\vert S^c}^1+ \lambda_2\bm{Q}_{S\vert
        S^c}^2\big)\big)\bm{1} \geq 0,
    \end{equation*}
    implying that
    \begin{align*}
        \tfrac{1}{2} \log\big(1+h\bm{1}^T\bm{Q}_{S\vert S^c}\bm{1}\big) 
        & \geq \tfrac{1}{2}\log\big(1+\lambda_1 h\bm{1}^T
        \bm{Q}^1_{S\vert S^c}\bm{1}
        +\lambda_2 h\bm{1}^T\bm{Q}^2_{S\vert S^c}\bm{1} \big) \\
        & \geq \lambda_1\tfrac{1}{2}\log\big(1+h\bm{1}^T
        \bm{Q}^1_{S\vert S^c}\bm{1}\big)
        +\lambda_2\tfrac{1}{2}\log\big(1+ h
        \bm{1}^T\bm{Q}^2_{S\vert S^c}\bm{1} \big).
    \end{align*}
    Thus $\psi_S(\bm{Q})$ is concave in $\bm{Q}$. Finally,
    \begin{align*}
        \min_{S\subset [N]} \big( & \tfrac{1}{2} \log(1+\card{S^c}g) 
        + \psi_S(\bm{Q}) \big) \\
        & \geq \min_{S\subset [N]} \Big(
        \lambda_1\big( \tfrac{1}{2}\log(1+\card{S^c}g) + \psi_S(\bm{Q}^1) \big) 
        +\lambda_2\big( \tfrac{1}{2}\log(1+\card{S^c}g) + \psi_S(\bm{Q}^2) \big)
        \Big) \\
        & \geq 
        \lambda_1 \min_{S\subset [N]} 
        \big( \tfrac{1}{2}\log(1+\card{S^c}g) + \psi_S(\bm{Q}^1) \big) 
        +\lambda_2 \min_{S\subset [N]}
        \big( \tfrac{1}{2}\log(1+\card{S^c}g) + \psi_S(\bm{Q}^2) \big),
    \end{align*}
    and hence $\psi(\bm{Q})$ is also concave in $\bm{Q}$. 

    Fix $\varepsilon > 0$, and assume that $\bm{Q}^\star$ achieves
    $\varepsilon$-optimality, i.e., $\bm{Q}^\star\geq 0$,
    $q^\star_{n,n}\leq 1$ for all $n\in[N]$, and 
    \begin{equation*}
        \psi(\bm{Q}^\star) \geq
        \sup_{\substack{\bm{Q}\geq 0:\\ q_{n,n}\leq 1 \forall n\in[N]}}
        \psi(\bm{Q})-\varepsilon.
    \end{equation*}
    Set
    \begin{equation*}
        \bm{Q} 
        = \frac{1}{N!}\sum_{\bm{P}}\bm{P}^T\bm{Q}^\star\bm{P},
    \end{equation*}
    where the sum is over all $N!$ permutation matrices on $[N]$. 
    
    Note that $\bm{Q}$ is positive semi-definite and satisfies
    $q_{n,n}\leq 1$ for all $n\in [N]$. Moreover, using the concavity
    and invariance under permutation of $\psi(\cdot)$, we obtain
    \begin{align*}
        \psi(\bm{Q})
        & \geq \frac{1}{N!}\sum_{\bm{P}}\psi\big(\bm{P}^T\bm{Q}^\star\bm{P}\big) \\
        & = \psi(\bm{Q}^\star),
    \end{align*}
    and hence $\bm{Q}$ is also an $\varepsilon$-optimal covariance matrix.
    Note that this $\bm{Q}$ has the form
    \begin{equation*}
        \rho \bm{1}_{N,N} + \kappa \bm{I}_{N},
    \end{equation*}
    for $\kappa\leq 1-\rho]$, and thus we can restrict the maximization
    of $\psi(\bm{Q})$ to matrices of this form.  Since the generalized
    Schur complement is monotonically increasing over the set of
    positive semi-definite matrices \cite[Theorem 3.1]{li00}, we can
    further restrict the value of $\kappa$ to be $1-\rho$. Denote the
    resulting matrix by $\bm{Q}^{\rho}$, i.e.,
    \begin{equation*}
        \bm{Q}^{\rho} \defeq 
        \begin{pmatrix}
            1 &\rho &\rho &\ldots &\rho \\
            \rho &1  &\rho &\ldots &\rho \\
            \rho & \rho & 1 &\ldots &\rho \\
            \vdots &\vdots  &\vdots &\ddots &\vdots \\
            \rho &\rho &\rho &\ldots &1 
        \end{pmatrix}.
    \end{equation*}
    Note that $\bm{Q}^\rho$ is positive semi-definite only if
    $\rho\in[-1/(N-1),1]$ (since otherwise the eigenvalue corresponding
    to the eigenvector $\bm{1}$ is negative).

    The upper bound on capacity in \eqref{eq:converse7} can thus be
    simplified to
    \begin{align}
        \label{eq:converse8}
        C & \leq \sup_{\rho\in[-1/(N-1),1]}
        \min_{S\subset [N]} 
        \big(\tfrac{1}{2}\log(1+\card{S^c}g) + \tfrac{1}{2}\log\big(1+ h\bm{1}^T
        \bm{Q}^\rho_{S\vert S^c} \bm{1}\big) \big) \nonumber\\
        & = \sup_{\rho\in[-1/(N-1),1]} \min_{n\in \{0,\ldots,N\}} 
        \big(\tfrac{1}{2}\log(1+(N-n)g) 
        + \tfrac{1}{2}\log\big(1+ h\bm{1}^T \bm{Q}^\rho_{[n]\vert [n]^c}
        \bm{1}\big) \big),
    \end{align}
    where $[0]$ is understood as the empty set and
    $[0]^c\defeq [N]$. Observe that the minimization in \eqref{eq:converse8}
    is over integers $n\in\{0,\ldots,N\}$ as opposed to subsets
    $S\subset[N]$ due to the symmetry in $\bm{Q}^\rho$. Note
    furthermore that instead of maximizing over arbitrary input
    distributions, we only have to maximize over the single real
    number $\rho$.

    We now compute the expression in parentheses in \eqref{eq:converse8}
    analytically. To this end, we need to compute $\bm{Q}^\rho_{[n]\vert
    [n]^c}$, which, by \eqref{eq:moore-penrose}, involves the
    computation of the generalized inverse
    $(\bm{Q}_{[n]^c,[n]^c}^\rho)^-$. We will first consider the case
    when $\bm{Q}_{[n]^c,[n]^c}^\rho$ is invertible, and then consider
    the remaining cases in which $\bm{Q}_{[n]^c,[n]^c}^\rho$ is not
    invertible. If $n\in \{1,\ldots,N-1\}$ and $\rho\in[-1/(N-1),1)$,
    then $\bm{Q}_{[n]^c,[n]^c}^\rho$ is invertible, and after some
    algebra, we obtain
    \begin{align}
        \label{eq:converse9a}
        \bm{1}^T\bm{Q}^\rho_{[n]\vert [n]^c} \bm{1}
        & = n\Big(1+(n-1)\rho-\frac{n(N-n)\rho^2}{1+(N-n-1)\rho}\Big).
    \end{align}

    We now consider the remaining cases, in which
    $\bm{Q}_{[n]^c,[n]^c}^\rho$ is not invertible.  If $\rho = 1$ and
    $n\in\{1,\ldots, N-1\}$, then
    \begin{equation}
        \label{eq:converse9b}
        \bm{1}^T\bm{Q}^1_{[n]\vert [n]^c} \bm{1}
        = 0.
    \end{equation}
    If $n=0$, then
    \begin{equation}
        \label{eq:converse9c}
        \bm{1}^T\bm{Q}^\rho_{[0]\vert [0]^c} \bm{1}
        = 0,
    \end{equation}
    and if $n=N$, then
    \begin{align}
        \label{eq:converse9d}
        \bm{1}^T\bm{Q}^\rho_{[N]\vert [N]^c} \bm{1}
        = \bm{1}^T\bm{Q}^\rho\bm{1} 
        = N(1+(N-1)\rho),
    \end{align}
    both for any $\rho\in[-1/(N-1),1]$.

    Denote by 
    \begin{equation*}
        \eta(\rho,n) 
        \defeq n\Big(1+(n-1)\rho-\frac{n(N-n)\rho^2}{1+(N-n-1)\rho}\Big)
    \end{equation*}
    the right-hand side of \eqref{eq:converse9a}.
    Note that $\eta(\rho,n)$ is well defined for all
    $\rho\in[-1/(N-1),1]$, $n\in\{0,\ldots,N\}$ except for $\eta(1,N)$
    and $\eta(-1/(N-1),0)$ (for which the expression involves dividing
    zero by zero). Moreover, from
    \eqref{eq:converse9a}--\eqref{eq:converse9d} we see that
    whenever $\eta(\rho,n)$ is well defined, it is equal to
    $\bm{1}^T\bm{Q}^\rho_{[n]\vert [n]^c} \bm{1}$. For the two cases in
    which $\eta(\rho,n)$ is not well defined, we have from
    \eqref{eq:converse9b}--\eqref{eq:converse9d} that
    for any $n\in[N]$, 
    \begin{subequations}
        \label{eq:limit}
        \begin{align}
            \lim_{\rho\uparrow 1} \eta(\rho,n)
            & = \bm{1}^T\bm{Q}^1_{[n]\vert [n]^c} \bm{1}, \\
            \lim_{\rho\downarrow -1/(N-1)} \eta(\rho,n)
            & = \bm{1}^T\bm{Q}^{-1/(N-1)}_{[n]\vert [n]^c} \bm{1},
        \end{align}
    \end{subequations}
    and in particular this holds for $n=0$ and $n=N$.  Thus we can write
    $\bm{1}^T\bm{Q}^{\rho}_{[n]\vert [n]^c} \bm{1}$ compactly as a
    function of $\eta(\rho,n)$ for any $n\in [N]$ and do not need to
    consider the boundary cases $n=0$, $n=N$, and $\rho=-1/(N-1)$,
    $\rho=1$ separately. Substituting \eqref{eq:converse9a} and
    \eqref{eq:limit} into \eqref{eq:converse8}, we obtain
    \begin{equation}
        \label{eq:converse10}
        C \leq \sup_{\rho\in(-1/(N-1),1)} \min_{n\in \{0,\ldots,N\}}
        \bigg(\tfrac{1}{2}\log(1+(N-n)g)  
        + \tfrac{1}{2}\log\Big(1+ n\Big(1+(n-1)\rho-\frac{n(N-n)\rho^2}{1+(N-n-1)\rho}\Big)h
        \Big)
        \bigg).
    \end{equation}
    Observe that the supremum in \eqref{eq:converse10} is only over
    $\rho\in(-1/(N-1),1)$ as opposed to $\rho\in[-1/(N-1),1]$ as in
    \eqref{eq:converse8}.

    We finally argue that the supremum can be restricted to be over
    values $\rho\in[0,1)$. Consider the derivative with respect to
    $\rho$ of the multiplier of the $h$ term in \eqref{eq:converse10},
    \begin{equation*}
        \frac{d}{d\rho}
        n\Big(1+(n-1)\rho-\frac{n(N-n)\rho^2}{1+(N-n-1)\rho}\Big) 
        = n\Big((n-1)-n(N-n)\rho\frac{2+(N-n-1)\rho}{\big(1+(N-n-1)\rho\big)^2}\Big).
    \end{equation*}
    If $\rho\in(-1/(N-1),0)$, then this derivative is non-negative, and
    thus the multiplier of $h$ in \eqref{eq:converse10} is non-decreasing
    in that range of $\rho$.  Since this is true simultaneously for all
    $n\in \{0,\ldots,N\}$, we can restrict the supremum to be 
    over the range $\rho\in[0,1)$. This proves the lemma.
\end{IEEEproof}

We now proceed to the proof of Theorem~\ref{thm:converse}.
As before, we denote the capacity of the diamond network by
\begin{equation*}
    C\defeq C(N,g,h).
\end{equation*}
We again consider the cases $\max\{g,Nh\} \geq 1$ and $\max\{g,Nh\} < 1$
separately.  Assume first that $\max\{g,Nh\} \geq 1$.  
Capacity is upper bounded by the minimum of the simple broadcast and
multiple-access cuts
\begin{align}
    \label{eq:proof_mult1}
    C 
    & \leq \min\big\{
    \tfrac{1}{2}\log(1+Ng),\tfrac{1}{2}\log(1+ N^2h)\big\} \nonumber\\
    & = \tfrac{1}{2}\log\big(1+N\min\{g,Nh\}\big).
\end{align}
Observe that \eqref{eq:proof_mult1} is valid regardless of the value of
$\max\{g,Nh\}$. 

Assume in the following that $\max\{g,Nh\} < 1$.  As before, we treat
the cases $g\leq h$, $g\in(h,N^2h)$, and $g\geq N^2h$ separately.
Consider first $g \leq h$. Using the upper bound in
\eqref{eq:proof_mult1}, we obtain
\begin{equation*}
    C 
    \leq \tfrac{1}{2}\log\big(1+Ng\big).
\end{equation*}

Consider then $g\in(h,N^2h)$. If $N\sqrt{gh} \geq 1$, then the
simplified form \eqref{eq:cutset_simple} of Lemma~\ref{thm:cutset} with
$N-n=\ceil{N^2h}\in\{0,\ldots,N\}$ (since $Nh\leq 1$)  yields
\begin{align*}
    C 
    & \leq \tfrac{1}{2}\log(1+(N-n)g) 
    + \tfrac{1}{2}\log\Big(1+ \frac{N^2}{N-n}h\Big) \\
    & = \tfrac{1}{2}\log(1+\ceil{N^2h} g) 
    + \tfrac{1}{2}\log\bigg(1+ \frac{N^2}{\ceil{N^2h}} h\bigg) \\
    & \leq \tfrac{1}{2}\log(1+g+N^2gh) + \tfrac{1}{2} \\
    & \leq \tfrac{1}{2}\log(1+2N^2gh) + \tfrac{1}{2},
\end{align*}
where we have used that $g\leq N^2gh$ since 
\begin{equation*}
    \textstyle
    N^2h \geq \sqrt{\smash[b]{g}\vphantom{N^2}}\sqrt{N^2h}
    = N\sqrt{\smash[b]{g}h\vphantom{N^2}}
    \geq 1.
\end{equation*}

Still assuming $g\in (h,N^2h)$, if $N\sqrt{gh} < 1$, then the simplified
form \eqref{eq:cutset_simple} of Lemma~\ref{thm:cutset} with
$N-n=\lceil N\sqrt{h\mbox{\small $/$}g}\,\rceil\in\{0,\ldots,N\}$
(since $g\geq h$ and hence $\sqrt{h\mbox{\small $/$}g}\leq 1$) shows that
\begin{align*}
    C 
    & \leq \tfrac{1}{2}\log(1+(N-n)g) 
    + \tfrac{1}{2}\log\Big(1+ \frac{N^2}{N-n}h\Big) \\
    & = \tfrac{1}{2}\log\big(1+{\textstyle \lceil N \sqrt{h\mbox{\small $/$}g}\,\rceil} g\big) 
    + \tfrac{1}{2}\log\bigg(1+ \frac{N^2}{\textstyle \lceil N\sqrt{h\mbox{\small $/$}g}\,\rceil} h\bigg) \\
    & \leq \tfrac{1}{2}\log\big(1+g+N{\textstyle\sqrt{gh}}\big) 
    + \tfrac{1}{2}\log\big(1+ N{\textstyle\sqrt{gh}}\big) \\
    & \leq \log\big(1+2N{\textstyle\sqrt{gh}}\big),
\end{align*}
where we have used that 
\begin{equation*}
    \textstyle 
    g \leq \sqrt{\smash[b]{g}\vphantom{N^2h}}\sqrt{N^2h} 
    = N\sqrt{\smash[b]{g}h\vphantom{N^2}}.
\end{equation*}

Finally, consider $g \geq N^2h$. The upper bound \eqref{eq:proof_mult1}
yields
\begin{equation*}
    C 
    \leq \tfrac{1}{2}\log\big(1+N^2h\big),
\end{equation*}
concluding the proof. \hfill\IEEEQED

\subsection{Proof of Corollary~\ref{thm:approx} (Capacity Approximation for Symmetric Networks)}
\label{sec:proofs_approx}

The corollary follows directly from Theorems~\ref{thm:achievability} and
\ref{thm:converse} using the inequalities
\begin{equation*}
    \log(1+ax)
    \begin{cases}
        \geq a\log(1+x), & \text{for $a\in[0,1], x\geq 0$} \\
        \leq a\log(1+x), & \text{for $a \geq 1, x\geq 0$},
    \end{cases}
\end{equation*}
and
\begin{equation*}
    \log(1+ax)
    \begin{cases}
        \geq \log(a)+\log(1+x), & \text{for $a\in[0,1], x\geq 0$} \\
        \leq \log(a)+\log(1+x), & \text{for $a \geq 1, x\geq 0$}.
    \end{cases}
\end{equation*}
\hfill\IEEEQED

\subsection{Proof of Theorem~\ref{thm:asymmetric} (Capacity
Approximation for Asymmetric Networks)}
\label{sec:proofs_asymmetric}

The idea of the proof is as follows. Group the relays into classes such
that all relays in the same class have approximately the same channel
gains. We argue that the number of classes needed is on the order
$\Theta(\log^2(N))$. Choose one such class, and set the constants
$\alpha_n=0$ for all relays not in this class. This effectively reduces
the network to a (almost) symmetric one, which we have already analyzed
in the earlier parts of this paper. By maximizing over which class to
choose, we get the largest rate achievable in this manner. This yields a
lower bound on $R_\delta\big(N,(g_n),(h_n)\big)$. We then argue that
this approach is close to optimal, by showing that capacity
$C\big(N,(g_n),(h_n))$ is upper bounded by $\Theta(\log^4(N))$ times the
maximum of the capacities of these classes. 

Recall the notation
\begin{equation*}
    [N] \defeq \{1,\ldots, N\}
\end{equation*}
and, for $S\subset[N]$, 
\begin{equation*}
    S^c \defeq [N]\setminus S.
\end{equation*}
Furthermore, in this section, we will use 
\begin{align*}
    g_S & \defeq (g_n)_{n\in S}, \\
    h_S & \defeq (h_n)_{n\in S}
\end{align*}
for $S\subset [N]$, and 
\begin{equation*}
    ah_S  \defeq (ah_n)_{n\in S}
\end{equation*}
for scalar $a\in\R$.

We want to partition $[N]$ into subsets such that for $n$ and
$\tilde{n}$ in the same subset the relays $v_n$ and $v_{\tilde{n}}$ have
approximately the same channel gains.  Moreover, we want the number of
required subsets to be small.  This is not directly possible if the
channel gains are very different. For example, consider $g_n=h_n=2^n$;
note, however, that in this case most of the relays are very weak
compared to the strongest one, and could hence be disregarded without
too much loss in rate. We formalize this idea by allowing some
``overload'' subsets (in the language of quantization theory) in the
partition of $[N]$, which correspond to relays that may have very
different channel gains, but that are all too weak to have much impact
on achievable rates.

Define 
\begin{align*}
    g^\star & \defeq \max_{n\in[N]} \min\{g_n, N^2 h_n\}, \\
    h^\star & \defeq \max_{n\in[N]} \min\{h_n, g_n\}.
\end{align*}
The quantities $g^\star$ and $h^\star$ are essentially the largest
channel gains, accounting for situations in which one of the channel
gains $g_n,h_n$ clearly dominates the other one. If we let 
$n$ be such that $h^\star = \min\{h_n, g_n\}$, then 
\begin{equation}
    \label{eq:asymmetric0a}
    g^\star \geq \min\{g_n, N^2 h_n\}
    \geq \min\{g_n, h_n\}
    = h^\star.
\end{equation}
Similarly, if $n$ is such that $g^\star = \min\{g_n, N^2 h_n\}$,
then
\begin{equation}
    \label{eq:asymmetric0b}
    h^\star \geq \min\{h_n, g_n\}
    \geq N^{-2} \min\{N^2h_n, g_n\}
    = N^{-2}g^\star.
\end{equation}
Thus, $g^\star$ and $h^\star$ can not be too different.

We are now ready to introduce the partition of $[N]$ mentioned
above. We start with the ``overload'' subsets. Define the sets
\begin{align*}
    T^1 
    & \defeq \big\{n\in [N]:
    g_n \leq N^{-3}g^\star \big\}, \\
    T^2 
    & \defeq \big\{n\in [N]\setminus T^1:
    h_n \leq N^{-3}h^\star \big\},
\end{align*}
i.e., $T^1$ and $T^2$ correspond to those relays that have channel
gains that are very weak compared to the strongest one in the
network. Set
\begin{equation*}
    L \defeq \floor{3\log(N)}.
\end{equation*}
For $\ell\in\{0,\ldots, L\}$, define
\begin{align*}
    T^1_{\ell} 
    & \defeq \big\{n\in [N]\setminus (T^1\cup T^2): 
    g_n\in(2^{-\ell-1}g^\star,2^{-\ell}g^\star], 
    h_n \geq g_n,
    \big\}, \\
    T^2_{\ell} 
    & \defeq \big\{n\in [N]\setminus (T^1\cup T^2\cup_{\tilde{\ell}} T^1_{\tilde{\ell}}): 
    g_n \geq N^2 h_n,
    h_n\in(2^{-\ell-1}h^\star,2^{-\ell}h^\star]
    \big\},
\end{align*}
i.e., $\{T^1_{\ell}\}$ and $\{T^2_{\ell}\}$ quantize those channel
gains for which one of $g_n,h_n$ dominates the other one.  Finally,
define for $k, \ell\in\{0,\ldots, L\}$,
\begin{align*}
    S_{k,\ell}
    \defeq \big\{ n \in [N]\setminus 
    \big(T^1\cup T^2\cup_{\tilde{\ell}}(T^1_{\tilde{\ell}}\cup T^2_{\tilde{\ell}})\big): 
    g_n\in(2^{-k-1}g^\star,2^{-k}g^\star], 
    h_n\in(2^{-\ell-1}h^\star,2^{-\ell}h^\star]
    \big\}.
\end{align*}
The subsets $\{S_{k,\ell}\}$ quantize the remaining channel
gains. The number of sets 
$T^1$, $T^2$, $\{T^1_\ell\}$, $\{T^2_\ell\}, \{S_{k,\ell}\}$
is equal to 
\begin{equation*}
    \tilde{L} \defeq (L+1)^2+2(L+1)+2 = \Theta(\log^2(N)).
\end{equation*}

We argue that $T^1$, $T^2$, $\{T^1_\ell\}$, $\{T^2_\ell\},
\{S_{k,\ell}\}$ partition $[N]$. The sets are clearly
disjoint, so we only need to show that their union covers $[N]$. If
either $g_n \leq N^{-3}g^\star$ or $h_n \leq N^{-3}h^\star$ then
$n\in T^1\cup T^2$. Assume in the following discussion that $g_n >
N^{-3}g^\star$ and $h_n > N^{-3}h^\star$. If $g_n \leq g^\star$ and
$h_n \leq h^\star$, then $n$ is an element of
$\{T^1_\ell\}$, $\{T^2_\ell\}$, or $\{S_{k,\ell}\}$. If $g_n >
g^\star$, then 
\begin{equation*}
    h_n \leq N^2h_n \leq g^\star\leq g_n,
\end{equation*}
so that $g_n \geq N^2h_n$ and $h_n=\min\{h_n,g_n\}\leq h^\star$. 
This implies that $n\in \cup_\ell T^2_\ell$. If $h_n > h^\star$, then
\begin{equation*}
    g_n \leq h^\star \leq h_n,
\end{equation*}
so that $h_n\geq g_n$ and $g_n = \min\{g_n, N^2h_n\} \leq
g^\star$.  This implies that $n\in \cup_\ell T^1_\ell$. Together,
this proves that we have properly partitioned $[N]$.

We are now ready for the proof of the upper bound on capacity. We
argue that the capacity of the diamond network with $N$ relays
cannot be much larger than the sum of the capacities of the
$\tilde{L}$ subchannels induced by the partition of $[N]$ defined
above. Formally, we argue that
\begin{align}
    \label{eq:asymmetric1a}
    C\big( N, g_{[N]}, h_{[N]}\big) 
    & \leq 
    C\big(\card{T^1}, g_{T^1},
    2\tilde{L}h_{T^1}\big)
    + C\big(\card{T^2}, g_{T^2},
    2h_{T^2}\big)  \nonumber\\
    & \quad +\sum_{i=1}^2\sum_{\ell=0}^L
    C\big(\card{T^i_{\ell}},
    g_{T^i_{\ell}},
    2\tilde{L}h_{T^i_{\ell}}\big) 
    +\sum_{k,\ell=0}^L
    C\big(\card{S_{k,\ell}},
    g_{S_{k,\ell}},
    2\tilde{L}h_{S_{k,\ell}}\big).
\end{align}
To see this, note that the right-hand side is the capacity of
$\tilde{L}$ parallel diamond networks each with unit input power
constraint. Moreover, increasing each channel gain $\sqrt{h_n}$ by a
factor of $\sqrt{2\tilde{L}}$ (or $\sqrt{2}$ in the case of $T^2$) is
equivalent to reducing the power of the additive noise at the
destination node of the parallel networks by a factor $1/(2\tilde{L})$
(or $1/2$ for $T^2$). We can now use these parallel networks 
to simulate the original $N$-relay diamond network by forcing the
input (at the source node $u$) to all the parallel networks to be
identical, and by summing up the outputs (at the destination node
$w$) of the parallel networks. This proves \eqref{eq:asymmetric1a}.

Next, we argue that the capacities of the asymmetric subnetworks in
\eqref{eq:asymmetric1a} can be upper bounded by the capacities of
symmetric diamond networks. Consider the subset
$S_{k,\ell}$. Since capacity is increasing in the channel
gains,
\begin{equation}
    \label{eq:asymmetric2a}
    C\big(\card{S_{k,\ell}},
    g_{S_{k,\ell}},
    2\tilde{L}h_{S_{k,\ell}}\big)
    \leq C\big(\card{S_{k,\ell}},
    2^{-k}g^\star,
    \tilde{L}2^{1-\ell}h^\star \big).
\end{equation}
Observe that the right-hand side is the capacity of a
\emph{symmetric} diamond network. Consider then $T^i_{\ell}$. By the same
argument
\begin{equation}
    \label{eq:asymmetric3a}
    C\big(\card{T^1_\ell},
    g_{T^1_\ell},
    2\tilde{L}h_{T^1_\ell}\big)
    \leq C\big(\card{T^1_\ell},
    2^{-\ell}g^\star,
    \infty\big), 
\end{equation}
and
\begin{equation}
    \label{eq:asymmetric4a}
    C\big(\card{T^2_\ell},
    g_{T^2_\ell},
    2\tilde{L}h_{T^2_\ell}\big)
    \leq C\big(\card{T^2_\ell},
    \infty,
    \tilde{L}2^{1-\ell}h^\star\big).
\end{equation}

It remains to consider $T^1$ and $T^2$. For the set $T^1$, we have
\begin{equation*}
    C\big(\card{T^1},
    g_{T^1},
    2\tilde{L}h_{T^1}\big)
    \leq C\big(N,
    N^{-3}g^\star,
    \infty\big).
\end{equation*}
From Theorem~\ref{thm:converse},
\begin{equation*}
    C\big(N, N^{-3}g^\star, \infty\big) 
    \leq \tfrac{1}{2}\log(1+N^{-2}g^\star).
\end{equation*}
By the definition of $g^\star$, there exists at least one
$n$ such that $g_n \geq g^\star$ and $h_n \geq N^{-2}g^\star$. Using
just this one relay $v_n$, a rate of at least
\begin{equation*}
    \tfrac{1}{2}\log(1+N^{-2}g^{\star})
\end{equation*}
is achievable.\footnote{This rate is achievable, for example, with
decode-and-forward. Note that we use decode-and-forward here only as
a proof technique to obtain the upper bound on capacity.
Achievability is based exclusively on (bursty) amplify-and-forward.}
For this $n$, we have
\begin{equation*}
    g_n \geq g^\star > N^{-3}g^\star,
\end{equation*}
and hence $n\notin T^1$. Moreover, using \eqref{eq:asymmetric0a},
\begin{equation*}
    h_n \geq N^{-2}g^\star \geq N^{-2}h^\star > N^{-3}h^\star,
\end{equation*}
and hence $n\notin T^2$. This $n$ is therefore an element of one of the
subsets $\{T^1_\ell\}$, $\{T^2_\ell\}$, $\{S_{k,\ell}\}$, and we obtain
from \eqref{eq:asymmetric2a}--\eqref{eq:asymmetric4a},
\begin{align}
    \label{eq:asymmetric5a}
    C\big(\card{T^1},g_{T^1}, \tilde{L}h_{T^1}\big) 
    \leq \max\bigg\{  \max_{\ell\in\{0,\ldots,L\}} 
    & C\big(\card{T^1_\ell}, 2^{-\ell}g^\star, \infty\big), 
    \max_{\ell\in\{0,\ldots,L\}}
    C\big(\card{T^2_\ell}, \infty, \tilde{L}2^{1-\ell}h^\star\big), \nonumber\\
    \max_{k,\ell\in\{0,\ldots,L\}}
    & C\big(\card{S_{k,\ell}},
    2^{-k}g^\star,
    \tilde{L}2^{1-\ell}h^\star \big)
    \bigg\}.
\end{align}

Similarly,
\begin{align*}
    C\big(\card{T^2},
    g_{T^2},
    2h_{T^2}\big)
    & \leq C\big(N,
    \infty,
    2N^{-3}h^\star\big) \\
    & \leq \tfrac{1}{2}\log(1+2N^{-1}h^\star) \\
    & \leq \tfrac{1}{2}\log(1+h^\star).
\end{align*}
By the definition of $h^\star$ there exists at least one $n$ such
that $h_n \geq h^\star$ and $g_n\geq h^\star$. Using this relay
$v_n$ alone, we achieve at least a rate of
\begin{equation*}
    \tfrac{1}{2}\log(1+h^{\star}).
\end{equation*}
For this $n$,
\begin{equation*}
    h_n \geq h^\star > N^{-3}h^\star,
\end{equation*}
and hence $n\notin T^2$. Moreover, using \eqref{eq:asymmetric0b}, 
\begin{equation*}
    g_n \geq h^\star \geq N^{-2}g^\star > N^{-3}g^\star,
\end{equation*}
and hence $n\notin T^1$. This $n$ is therefore an element of one of the
subsets $\{T^1_\ell\}$, $\{T^2_\ell\}$, $\{S_{k,\ell}\}$, and we obtain
again from \eqref{eq:asymmetric2a}--\eqref{eq:asymmetric4a},
\begin{align}
    \label{eq:asymmetric6a}
    C\big(\card{T^2}, g_{T^2}, 2h_{T^2}\big) 
    \leq \max\bigg\{ \max_{\ell\in\{0,\ldots,L\}} 
    & C\big(\card{T^1_\ell}, 2^{-\ell}g^\star, \infty\big), 
    \max_{\ell\in\{0,\ldots,L\}} 
    C\big(\card{T^2_\ell}, \infty, \tilde{L}2^{1-\ell}h^\star\big), \nonumber\\
    \max_{k,\ell\in\{0,\ldots,L\}}
    & C\big(\card{S_{k,\ell}},
    2^{-k}g^\star,
    \tilde{L}2^{1-\ell}h^\star \big)
    \bigg\}.
\end{align}

Substituting \eqref{eq:asymmetric2a}--\eqref{eq:asymmetric6a} into
\eqref{eq:asymmetric1a}, we obtain 
\begin{align}
    \label{eq:asymmetric7a}
    C\big(N, g_{[N]}, h_{[N]} \big)
    \leq \tilde{L} \max\bigg\{
    \max_{\ell\in\{0,\ldots,L\}} 
    & C\big(\card{T^1_\ell}, 2^{-\ell}g^\star, \infty\big), 
    \max_{\ell\in\{0,\ldots,L\}}  
    C\big(\card{T^2_\ell}, \infty, \tilde{L}2^{1-\ell}h^\star\big), \nonumber\\
    \max_{k,\ell\in\{0,\ldots,L\}}
    & C\big(\card{S_{k,\ell}},
    2^{-k}g^\star,
    \tilde{L}2^{1-\ell}h^\star \big)
    \bigg\}.
\end{align}
This concludes the proof of the upper bound on capacity.

We continue with the proof of achievability. Fix
$k,\ell\in\{0,\ldots, L\}$, and recall that $\alpha_n$ is the
constant determining the amplification at relay $v_n$. 
Assume we set $\alpha_n=0$ for all $n\notin S_{k,\ell}$. This
results in a network in which all but the relays in
$S_{k,\ell}$ are removed. Thus
\begin{equation}
    \label{eq:asymmetric1b}
    R_{\delta}\big(N,g_{[N]},h_{[N]}\big)
    \geq R_{\delta}\big(\card{S_{k,\ell}},
    g_{S_{k,\ell}},
    h_{S_{k,\ell}}\big).
\end{equation}
Moreover, since $R_\delta$ is increasing in the channel gains,
\begin{equation}
    \label{eq:asymmetric2b}
    R_{\delta}\big(\card{S_{k,\ell}},
    g_{S_{k,\ell}},
    h_{S_{k,\ell}}\big)
    \geq R_{\delta}\big(\card{S_{k,\ell}},
    2^{-k-1}g^\star,
    2^{-\ell-1}h^\star\big).
\end{equation}

With this, we have lower bounded the rate achievable for the asymmetric
diamond network by the one of a symmetric diamond network (with fewer
relays and smaller channel gains).  We can thus apply the results from
Section~\ref{sec:main_symmetric} to obtain
\begin{equation}
    \label{eq:asymmetric3b}
    \sup_{\delta\in (0,1]} R_{\delta}\big(\card{S_{k,\ell}},
    2^{-k-1}g^\star,
    2^{-\ell-1}h^\star\big) 
    \geq
    \frac{1}{112\tilde{L}}
    C \big(\card{S_{k,\ell}},
    2^{-k}g^\star,
    \tilde{L}2^{1-\ell}h^\star\big),
\end{equation}
where the factor $1/(112\tilde{L})=1/(8\tilde{L}\times 14)$ is composed
of a factor $8\tilde{L}$ to offset the increase of the channel gains
to the relay by two and the increase of the channel gains from
the relays by $4\tilde{L}$ (see Theorem~\ref{thm:achievability}) and
of a factor $14$ to go from rate achievable with bursty amplify-and-forward
to capacity (see Theorem~\ref{thm:converse} and
Corollary~\ref{thm:approx}).

Combining \eqref{eq:asymmetric1b}, \eqref{eq:asymmetric2b}, and
\eqref{eq:asymmetric3b} yields
\begin{equation}
    \label{eq:asymmetric4b}
    \sup_{\delta\in(0,1]}R_{\delta}\big(N,g_{[N]},h_{[N]}\big) \\
    \geq
    \frac{1}{112\tilde{L}}
    C \big(\card{S_{k,\ell}},
    2^{-k}g^\star,
    \tilde{L}2^{1-\ell}h^\star\big).
\end{equation}
A similar argument, setting $\alpha_n=0$ for $n$ outside $T^i_\ell$,
shows that
\begin{align}
    \label{eq:asymmetric5b}
    \sup_{\delta\in(0,1]}
    R_{\delta} \big(N,g_{[N]},h_{[N]}\big) 
    & \geq \sup_{\delta\in(0,1]}R_{\delta}\big(\card{T^1_\ell},
    g_{T^1_\ell}, 
    h_{T^1_\ell} \big), \nonumber\\
    & \geq \sup_{\delta\in(0,1]}R_{\delta}\big(\card{T^1_\ell},
    2^{-\ell-1}g^\star, 
    2^{-\ell-1}g^\star
    \big), \nonumber\\
    & \geq \frac{1}{28}
    C\big(\card{T^1_\ell}, 2^{-\ell}g^\star, \infty \big),
\end{align}
and
\begin{align}
    \label{eq:asymmetric6b}
    \sup_{\delta\in(0,1]}
    R_{\delta} \big(N,g_{[N]},h_{[N]}\big) 
    & \geq \sup_{\delta\in(0,1]}R_{\delta}\big(\card{T^2_\ell},
    g_{T^2_\ell}, 
    h_{T^2_\ell} \big), \nonumber\\
    & \geq \sup_{\delta\in(0,1]}R_{\delta}\big(\card{T^2_\ell},
    N^2 2^{-\ell-1}h^\star, 
    2^{-\ell-1}h^\star
    \big), \nonumber\\
    & \geq \frac{1}{56\tilde{L}}
    C\big(\card{T^1_\ell}, \infty, \tilde{L}2^{1-\ell}h^\star\big),
\end{align}
for all $\ell\in\{0,\ldots, L\}$.

We can optimize over the lower bounds in \eqref{eq:asymmetric4b},
\eqref{eq:asymmetric5b}, and \eqref{eq:asymmetric6b} to obtain
\begin{align*}
    \sup_{\delta\in(0,1]}R_{\delta} \big(N,g_{[N]},h_{[N]}\big) 
    \geq \frac{1}{112\tilde{L}}
    \max\bigg\{
    \max_{\ell\in\{0,\ldots,L\}} 
    & C\big(\card{T^1_\ell}, 2^{-\ell}g^\star, \infty\big),
    \max_{\ell\in\{0,\ldots,L\}} 
    C\big(\card{T^2_\ell}, \infty, \tilde{L}2^{1-\ell}h^\star\big),
    \nonumber\\
    \max_{k,\ell\in\{0,\ldots,L\}}
    & C\big(\card{S_{k,\ell}},
    2^{-k}g^\star,
    \tilde{L}2^{1-\ell}h^\star \big)
    \bigg\}.
\end{align*}
Comparing this with the upper bound \eqref{eq:asymmetric7a} shows that
\begin{equation*}
    C(N,g_{[N]},h_{[N]})
    \leq 112\tilde{L}^2\sup_{\delta\in(0,1]}R_\delta(N,g_{[N]},h_{[N]}).
\end{equation*}
Using that
\begin{equation*}
    \tilde{L} \leq (3\log(N)+1)^2+6\log(N)+4
\end{equation*}
shows that there exists a universal constant $K< \infty$ (and, in
particular, independent of $g_{[N]}$, $h_{[N]}$, and
$N$) such that $112\tilde{L}^2 \leq K\log^4(N)$ for $N\geq 2$. This
concludes the proof of the theorem.  \hfill\IEEEQED

\section{Conclusion}
\label{sec:conclusion}

We presented an approximation of the capacity of the symmetric Gaussian
$N$-relay diamond network. The capacity was characterized up to a $1.8$
bit additive gap and a factor $14$ multiplicative gap uniformly for all
channel gains and number of relays. The inner bound in this approximate
characterization relies on bursty amplify-and-forward, showing that this
scheme is good simultaneously at low and high rates, uniformly in the
channel gains and in the number of relays $N$. The upper bound resulted
from a careful evaluation of the cut-set bound. We argued that all $2^N$
possible cuts in the diamond network need to be evaluated
simultaneously, and that the standard approach of only considering the
minimum of the broadcast and multiple-access cuts is insufficient to
derive uniform capacity approximations. We extended this approach to
asymmetric diamond networks, for which we showed that bursty
amplify-and-forward achieves capacity up to a multiplicative gap of a
factor $O(\log^4(N))$ with pre-constant in the order notation
independent of the channel gains.

The results in this paper show that, at least for symmetric diamond
networks, it is possible to derive capacity approximations that are
independent of the network size. Deriving such uniform capacity
approximations for general networks remains an open problem.

\bibliography{journal_abbr,diamond}

\begin{thebibliography}{10}

\bibitem{schein00}
B.~Schein and R.~Gallager, ``The {G}aussian parallel relay network,'' in {\em
  Proc. IEEE ISIT}, p.~22, June 2000.

\bibitem{schein01}
B.~Schein, {\em Distributed Coordination in Network Information Theory}.
\newblock PhD thesis, Massachusetts Institute of Technology, 2001.

\bibitem{gastpar05}
M.~Gastpar and M.~Vetterli, ``On the capacity of large {G}aussian relay
  networks,'' {\em IEEE Trans. Inf. Theory}, vol.~51, pp.~765--779, Mar. 2005.

\bibitem{kochman08}
Y.~Kochman, A.~Khina, U.~Erez, and R.~Zamir, ``Rematch and forward for parallel
  relay networks,'' in {\em Proc. IEEE ISIT}, pp.~767--771, July 2008.

\bibitem{rezaei09}
S.~S.~C. Rezaei, S.~O. Gharan, and A.~K. Khandani, ``A new achievable rate for
  the {G}aussian parallel relay channel,'' in {\em Proc. IEEE ISIT},
  pp.~194--198, June 2009.

\bibitem{xue07}
F.~Xue and S.~Sandhu, ``Cooperation in a half-duplex {G}aussian diamond relay
  channel,'' {\em IEEE Trans. Inf. Theory}, vol.~53, pp.~3806--3814, Oct. 2007.

\bibitem{bagheri09}
H.~Bagheri, A.~S. Motahari, and A.~K. Khandani, ``On the capacity of the
  half-duplex diamond channel,'' {\em arXiv:0911.1426 [cs.IT]}, Nov. 2009.
\newblock submitted to IEEE Transactions on Information Theory.

\bibitem{kang11}
W.~Kang and S.~Ulukus, ``Capacity of a class of diamond channels,'' {\em IEEE
  Trans. Inf. Theory}, vol.~57, pp.~4955--4960, Aug. 2011.

\bibitem{etkin08}
R.~Etkin and D.~N.~C. Tse, ``{G}aussian interference channel capacity to within
  one bit,'' {\em IEEE Trans. Inf. Theory}, vol.~54, pp.~5534--5562, Dec. 2008.

\bibitem{avestimehr09}
A.~S. Avestimehr, S.~N. Diggavi, and D.~N.~C. Tse, ``Wireless network
  information flow: A deterministic approach,'' {\em IEEE Trans. Inf. Theory},
  Apr. 2011.

\bibitem{ozgur10c}
A.~{\"O}zg{\"u}r and S.~Diggavi, ``Approximately achieving {G}aussian relay
  network capacity with lattice codes,'' in {\em Proc. IEEE ISIT},
  pp.~669--673, June 2010.
\newblock See also arXiv:1005.1284 [cs.IT].

\bibitem{lim11}
S.~H. Lim, Y.-H. Kim, A.~E. Gamal, and S.-Y. Chung, ``Noisy network coding,''
  {\em IEEE Trans. Inf. Theory}, vol.~57, pp.~3132--3152, May 2011.

\bibitem{gupta00a}
P.~Gupta and P.~R. Kumar, ``The capacity of wireless networks,'' {\em IEEE
  Trans. Inf. Theory}, vol.~46, pp.~388--404, Mar. 2000.

\bibitem{ozgur07b}
A.~{\"O}zg{\"u}r, O.~L{\'e}v{\^e}que, and D.~N.~C. Tse, ``Hierarchical
  cooperation achieves optimal capacity scaling in ad hoc networks,'' {\em IEEE
  Trans. Inf. Theory}, vol.~53, pp.~3549--3572, Oct. 2007.

\bibitem{ghaderi09}
J.~Ghaderi, L.-L. Xie, and X.~Shen, ``Hierarchical cooperation in ad hoc
  networks: Optimal clustering and achievable throughput,'' {\em IEEE Trans.
  Inf. Theory}, vol.~55, pp.~3425--3436, Aug. 2009.

\bibitem{niesen09a}
U.~Niesen, P.~Gupta, and D.~Shah, ``On capacity scaling in arbitrary wireless
  networks,'' {\em IEEE Trans. Inf. Theory}, vol.~56, pp.~3959--3982, Sept.
  2009.

\bibitem{niesen09c}
U.~Niesen, ``Interference alignment in dense wireless networks,'' {\em IEEE
  Trans. Inf. Theory}, vol.~57, pp.~2889--2901, May 2011.

\bibitem{cover91}
T.~M. Cover and J.~A. Thomas, {\em Elements of Information Theory}.
\newblock Wiley, 1991.

\bibitem{muirhead82}
R.~J. Muirhead, {\em Aspects of Multivariate Statistical Theory}.
\newblock Wiley, 1982.

\bibitem{thomas87}
J.~A. Thomas, ``Feedback can at most double {G}aussian multiple access channel
  capacity,'' {\em IEEE Trans. Inf. Theory}, vol.~33, pp.~711--716, Sept. 1987.

\bibitem{li00}
C.-K. Li and R.~Mathias, ``Extremal characterizations of the {S}chur complement
  and resulting inequalities,'' {\em SIAM Review}, vol.~42, pp.~233--246, June
  2000.

\bibitem{marshall79}
A.~W. Marshall and I.~Olkin, {\em Inequalities: Theory of Majorization and Its
  Applications}.
\newblock Academic Press, 1979.

\end{thebibliography}

\end{document}